# COVENTRY UNIVERSITY

Faculty of Engineering, Environment and Computing

School of Computing, Electronics and Mathematics

Reviewing the Scope and Impact of Implementing a Modernised IT Event-Driven Architecture from Traditional Architecture using Agile Frameworks – A Case study of Bi-modal operational strategy

Author: Sunday David Ubur

SID: 8460552

Supervisor: Simon Billings

Submitted in partial fulfilment of the requirements for the Degree of Master of Science in Software Development

Academic Year: 2019/20

# Declaration of Originality

I declare that this project is all my own work and has not been copied in part or in whole from any other source except where duly acknowledged. As such, all use of previously published work (from books, journals, magazines, internet etc.) has been acknowledged by citation within the main report to an item in the References or Bibliography lists. I also agree that an electronic copy of this project may be stored and used for the purposes of plagiarism prevention and detection.

# Statement of copyright

I acknowledge that the copyright of this project report, and any product developed as part of the project, belong to Coventry University. Support, including funding, is available to commercialise products and services developed by staff and students. Any revenue that is generated is split with the inventor/s of the product or service. For further information please see www.coventry.ac.uk/ipr or contact ipr@coventry.ac.uk.

# Statement of ethical engagement

I declare that a proposal for this project has been submitted to the Coventry University ethics monitoring website (https://ethics.coventry.ac.uk/) and that the application number is listed below (Note: Projects without an ethical application number will be rejected for marking)

Signed: 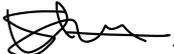   Date: 13th August 2020

Please complete all fields.

| First Name: | Sunday David |
|---|---|
| Last Name: | Ubur |
| Student ID number | 8460552 |
| Ethics Application Number | P107910 |
| 1st Supervisor Name | Dr Simon Billings |
| 2nd Supervisor Name | Dr Rochelle Sassman |

**This form must be completed, scanned and included with your project submission to Turnitin. Failure to append these declarations may result in your project being rejected for marking.**

# Abstract

Proposing and implementing software systems, especially web applications for e-commerce using the traditional monolithic approach has been the norm, however, as new user requirements force organisations and developers to add more functionalities to their systems, and as user demands increases, the performance of monolith applications decreases, and maintenance costs rise. These issues have necessitated a need for a better system and that is where Microservices event-driven applications come in. This thesis investigates the scope and impact of migrating to Microservices, using agile frameworks, and bi-modal IT strategy.  To obtain quantitative data for analysis, prototypes of monolith and microservice websites were integrated with Dropwizard, an open source metrics, and Apache JMeter, and response time and error rate readings were collected. Results showed that monoliths have a faster response time when the number of user request calls is within tolerant range, better than Microservices, but as complexity grows over time and there is increased user request calls, Microservices appears to perform better.

# Acknowledgements

*"If I have seen further, it is by standing upon the shoulders of giants"*
*(Isaac Newton, 1687)*

My academic journey and success have been due to the support and encouragement I received from those who believed in me and have given me the opportunity.

Firstly, I wish to acknowledge and thank the UK government's Chevening FCO for the Chevening Awards to study a Masters' degree in the UK. This is a rare and unique opportunity.

Secondly, would like to thank my supervisors' Dr Simon Billings and Dr Carey Pridgeon for their encouragement, challenges and insightful comments and recommendations.

Thirdly, I would like to thank the team at Salescache, especially my employer Mr Michael Orji for the opportunity to work with their ecommerce system, an opportunity that motivated the choice of this research topic for investigation.

Finally, my profound thanks to my lecturers for their support and cooperation throughout the duration of my studies at the university.

Thank you!!

# Table of Contents

# List of Figures

# List of Tables

# Abbreviations

SOA     Service Oriented Architecture

ESB      Enterprise Service Bus

MSA     Micro Signal Architecture

EDA     Event Driven Architecture

CEP      Complex Event Processing

WSDL   Web Services Business Process Execution Language

API      Application Programming Interface

REST    Representational State Transfer

HTTP    Hypertext Transfer Protocol

BI        Business Intelligence

EDM    Electronic Document Management

BPM    Business Process Management

BRMS   Business Rules Management System

# Chapter 1

# Introduction

Microservices is getting more popular in recent years, and many companies and agencies are migrating their monolith applications to microservices event-driven architecture. This architecture allows developers to independently develop and deploy services and ease the adoption of agile process. On the other hand, using agile frameworks with monolith architecture is very difficult as teams experience challenges collaborating on a single application at the same time. Ironically, many companies are still hesitant to migrate because they think microservices is a hype or because they are not aware of the migration process and the benefits and matters related to migration. For this reason, this research intends to build basic monolith and microservice prototypes and conduct experiments to measure their metrics that can be analysed to investigate the scope and impact of microservices compared to monoliths.

## 1.1 Problem Statement

Many business enterprises, companies and government agencies that adopt IT as critical part of their daily operations still depend on traditional or legacy architecture software although most of the components are becoming obsolete and incur about 80 percent of the IT department's operational costs (Newman, 2017). They find it difficult to apply modern IT practices such as Agile frameworks as it is difficult for a team of developers to work collaboratively on a traditional legacy or monolith architected software, thereby causing maintenance to become more difficult. Salescache, an online discounts shop is one example of several companies whose enterprise system was build using the monolith approach. It resulted in the system being tightly coupled; difficult to maintain, and the entire system needing to be rebuilt using the event-driven architecture framework. However, findings show that many companies are sceptical in migrating to the microservices event-driven, not because they don't embrace the new technology and the benefit agility offers, but because they don't understand how to migrate while ensuring little or no disruption to the old system (Purcell, 2017). There is also the problem of financial aspects of migrating to microservices as managed services in public cloud platforms have very different pricing models, and the choice of a microservice runtime or data storage solution can have a significant impact on the cost-effectiveness of running a microservice system in the cloud. These issues inspired this research work to attempt and investigate the efficiency of event-driven architecture compared to the traditional architecture, analyse the scope and impact of migrating, and the benefits agile adoption will have on such enterprises, companies and agencies.

## 1.2 Project Objectives

i.  A review of Traditional Legacy Architecture and cost of IT maintenance, upgrade and modernisation
ii. A review of Market adoption of Event-driven Architectures: A Case Study of Netflix
iii. Analysis of Agile Frameworks in the Transformation from Traditional Legacy Architecture to Event-driven Architecture
iv. Benefits of Agile Adoption in IT Modernisation when implementing Event-driven Architecture to replace Monolithic Systems
v.  Operationalisation of Event-driven architecture using Bi-Modal Strategy in decommissioning systems with Traditional Legacy Architecture

## 1.3 Purpose of the Study

The purpose of this study is to determine the scope and impact of migrating from traditional monolith architecture to microservices event-driven architecture by analysing the challenges of maintaining a monolith system, migrating to EDA while avoiding disruption, and the application of agile frameworks in the process. The study aims to meet the project objectives by answering the following research questions, labelled **RQ 1-2**.

**RQ 1**: How to migrate to microservices event-driven architecture while not disrupting ongoing business operations that depend on the traditional architecture system?

Intending to migrate while ensuring minimal or no disruption to current system is a challenge and this research question aims to clarify that using the bimodal-operation strategy via relevant literature and industrial application guide.

**RQ 2**: What are the scope and impact of migrating to microservices event-driven architecture, including using agile frameworks?

This research question will analyse and compare the cost of maintenance of both architectures and their performance characteristics. This will help organisations who are reluctant to migrate their systems from monolith to microservices to make informed choices.

### 1.4 Intended Users

  i. The academic community: At the end of this project, we expect to have more in-depth knowledge of various event-driven architectures currently available and determine the preferred architecture for enterprise systems.
 ii. Software developers aiming to use the event-driven architecture in their practices: Encourage painless use of agility as collaborating teams will be dealing with separate modules that are highly independent and loosely coupled.
iii. Salescache: the report may help the company know how to maintain their system going forward, as well as make the best choice of architecture in their future developments.
 iv. Career progression

# Chapter 2

# Literature Review

In this chapter researcher conducts a review of literature by focusing on the research objectives and research topic, that is "reviewing the scope and impact of implementing a modernised IT event-driven architecture from traditional architecture using agile frameworks – a case study of bi-modal operational strategy".

## 2.1 A Review of Event-Driven Architecture

### 2.1.1 Event Driven Architecture Style

In 2006, Gartner closed an analytic group that focused on Event-driven architecture (EDA) (Nkomo and Coetzee, 2019). The term EDA was proposed by Gartner Roy W. Schulte, an analyst three years earlier in the Growing Role of Events in Enterprise Applications. Maybe the concept was too revolutionary, or maybe the idea of Complex Event processing (CEP) diverted the main attention, but because of the delay in introducing technologies such as RFID, it went into the shadows, taking with it EDA, it doesn't matter. For us, something else is important: firstly, unlike Service-oriented architecture (SOA), EDA is really an architecture (Talbot et al., 2019), i.e. a specific approach to building an enterprise information system; secondly, the concept of event-driven architecture explains why today interest in the enterprise service bus (ESB) is ahead of interest in SOA.

To begin with, why SOA is not an architecture. SOA was architecture at the time of its appearance, while it was associated with such a picture (Yan et al., 2019). But the idea of pre-registering services in the registry and detecting them during execution did not work and the real service-oriented architecture looks more like this. But EDA is really architecture, i.e. a pattern that describes the interaction between software components, each of which has a specific role: one component generates a message, the other intercepts and processes it, if necessary, invokes third components for this. Generally speaking, this is what integration environments have always dealt with.

The next subtlety is that to build a service-oriented architecture, the Enterprise service bus, generally speaking, is not needed. Applications themselves can provide services (Talbot et al., 2019). Enterprise Service Bus (ESB) was positioned as a tool for developing services for legacy applications. But if such an application develops at least somehow, then you can remove the program interfaces from it and wrap them with services without the help of the ESB. And if the application is already completely inherited and no one understands how it functions there within itself, then it will not be possible to "pull out" services from it using the ESB. But to build an event-driven architecture, an integration environment is needed. Events must be intercepted, converted, routed, etc. Thus, EDA puts everything in its place (Simchev, 2018).

The event-driven architecture includes **event providers** that create event flows, and **event consumers** that listen for these events.

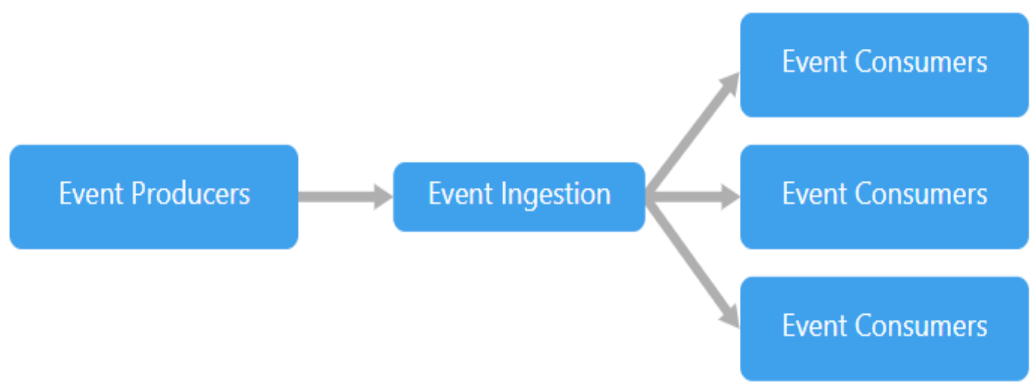

Figure 2.1: Event Driven Architecture Style (Dehbi, 2016)

Events are delivered almost instantly, which allows consumers to immediately respond to events. Suppliers are not affiliated with consumers - no supplier knows who is listening to his events. Consumers are also independent of each other, and each of them receives all the events (Tang et al., 2015). This is an important difference from the competing customer template , in which users retrieve messages from the queue and each message is processed only once (if no errors occur). In some systems, such as the Internet of things, events are processed in huge volumes.

In the above logical diagram, each type of consumer is indicated by a separate block. In practice, multiple instances of each supplier are usually used so that they do not become a single point of failure. Multiple instances may also be required to process events in the right volumes and / or at the right speed. In addition, each receiver can handle events in multiple threads. This can create problems if events need to be handled in order or require exactly one semantics.

## 2.2 A Review of Microservices Architecture

The term "Microservice Architecture" has been used in the past few years as a description of how to design applications as a set of independently deployed services (Kang, Le and Tao, 2016). While there is no exact description of this architectural style, there is a certain general set of characteristics: organisation of services around business needs, automatic deployment, transfer of logic from the message bus to receivers (endpoints) and decentralised control over languages and data.

According to (Theorin et al., 2015), monolithic server is a fairly obvious way to build such systems. All the logic for processing requests is performed in a single process, while you can use the capabilities of your programming language to divide the application into classes, functions, and namespace. One can run and test the application on the developer's machine and use the standard deployment process to check for changes before putting them into production. One can scale a monolithic application horizontally by "running multiple physical servers behind a load" balancer. Its roots go far into the past, at least to the design principles used in Unix. But we nevertheless believe that not enough people take this style into account and that many applications will benefit if they start using this style.

**2.2.1 Microservice Architecture Properties**

As mentioned by (Francesco, 2017), "there is a formal definition of the style of microservices, but we can try to describe what we consider to be common characteristics" of applications using this style. They are not always found in one application all at once, but, as a rule, each such application includes most of these characteristics.

Remote calls are slower than calls within the process, and therefore the Application Programming Interface (API) should be less detailed (coarse-grained), which often leads to inconvenience in use. In a first approximation, we can observe that services are related to processes as one to one. In fact, a service can contain "many processes that will always be developed and deployed together". For example, the application process and the database process that only this application uses.

The microservice community prefers an alternative approach: smart message receivers and transmission channels (Yarygina and Bagge, 2018). Applications built using the microservice architecture tend to be as decoupled and cohesive as possible: they contain their own domain logic and act more as filters in the classic Unix sense- they receive requests, apply logic and send an answer. Instead of complex protocols such as Web Services Business Process Execution Language (WS-BPEL), they use simple Representational State Transfer (REST) protocols. The two most commonly used protocols are Hypertext Transfer Protocol (HTTP) requests through the resource API and lightweight messaging.

Teams that practice "microservice architecture use the same principles and protocols that the World Wide Web" is built on (and essentially Unix). Frequently used resources can be cached with very little effort from developers or IT administrators. A second commonly used communication tool is a lightweight message bus. Such an infrastructure usually does not contain domain logic - simple implementations like RabbitMQ do nothing but provide an asynchronous factory. In this case, logic exists at the ends of this bus - in services that send and receive messages (Gannon, Barga and Sundaresan, 2017).

In a monolithic application, components work in one process and communicate with each other through method calls. The biggest problem in "changing the monolith to microservices lies in changing the communication template" (Strljic et al., 2019). One-to-one naive porting "leads to coupled communications" that don't work too well. There is the need to reduce the number of communications between modules.

Breaking a monolith into microservices event-driven architecture, we have a choice of how to build each of them (Mayer and Weinreich, 2018). Microservice teams also prefer a different standardisation approach, thus, instead of using a set of predefined standards written by someone, "they prefer the idea of building useful tools that other developers can use to solve similar problems" (Mayer and Weinreich, 2018). These tools are usually extracted from the code of one of the projects and shared between different teams, sometimes using the internal open source model. Now that "git and GitHub have become the de facto standard version control system, open source practices are becoming more and more popular in company internal projects" (Mayer and Weinreich, 2018).

Netflix is a good example of an organisation that follows this philosophy (Gannon, Barga and Sundaresan, 2017). Sharing useful and, moreover, library-tested code on battle servers encourages other developers to solve similar problems in a similar way, leaving the possibility of choosing a different approach if necessary. Shared libraries tend to focus on common issues related to data storage, inter process communication, and infrastructure automation.

**2.2.3 Microservices and EDA**

The "microservice style is very similar to what some proponents of EDA" are promoting (Strljic et al., 2019). The problem, however, is that the term EDA has too many different meanings and, as a rule, what people call "EDA" differs significantly from the style described here, usually due to the excessive focus on the ESB used for integrating monolithic applications. In particular, we have seen so many unsuccessful EDA implementations (starting with the tendency to hide complexity behind ESBs, ending with failed initiatives lasting several years that cost millions of monies and did not bring any benefit), that it is sometimes too difficult to ignore these problems.

These EDA issues have led some microservice proponents to abandon the term "EDA," while others consider microservices to be a form of EDA, or perhaps the correct implementation of EDA (Strljic et al., 2019). In any case, the fact that EDA has different meanings means that it is useful to have a separate term for this architectural style. Of course, many of the practices used in microservices come from the experience of integrating services in large organisations. The Tolerant Reader template is one example. Another example is the use of simple protocols introduced as a reaction to centralised standards, the complexity of which is breath-taking (Hasselbring and Steinacker, 2017).

2.3 A **Review of Event-Driven Architecture Using Microservices As Platform**

  2.3.1 **Event-Driven Microservices Patterns**

As mentioned earlier, the microservice architecture style is a method of building an application in the form of a discrete but independent management configuration based on explicit knowledge of the business (Perez et al., 2018). Microservice-based applications can be event driven. There are interesting engineering projects in rapid development and further development. As stated by (Monteiro et al., 2018), the microservice approach offers developers greater flexibility in choosing the perfect layout for information middleware. Each use case has its own specific requirements, which require various information innovations, such as Apache Kafka, RabbitMQ or even NoSQL event information matrices, such as Apache Geode / Pivotal GemFire.

In the Micro Signal Architecture (MSA) approach, a typical architectural project captures events using a nursery-only sequence of events, for example, Kafka or MapR feeds that Kafka updates. Using MapR Streams, events are grouped into coherent sets of events called subjects (Perez et al., 2018). Events are transmitted in the received request. Unlike the line, events are permanent, and after transmission they remain significant in a packet that can be accessed by different clients.

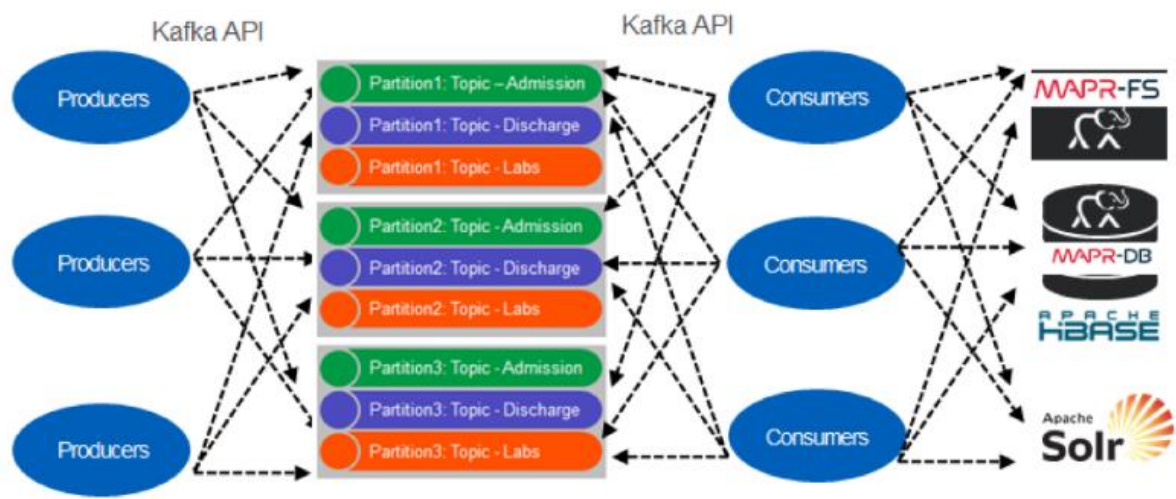

Figure 2.2: Kafka Based Architecture *(Perez et al., 2018)*

Event-Driven Architecture (EDA) is a popular, non-competitive distribution architecture project that solves the problems of dispersed information. It is exceptionally customisable and adaptable. In EDA, every microservice sends an event when something unusual happens (Khaddaj and Srour, 2016). For example, the request service dispatches a new event if a request has been made or modified. Various microservices buy events that interest them (for example, the inventory service will buy new request events, as this will reduce the inclusion of irrelevant products in the inventory database). The event can be called a "big change of state". Events can be used to conduct trading that covers various jurisdictions (Mayer and Weinreich, 2018). We can talk about exchanges with a sequence of steps, where each sequence is a micro-service that updates or creates a commercial substance and distributes an event that triggers the next phase. EDA has two main regions: intermediary and agent.

Event-driven microservices and architecture are the preferred methods for updating today's universal cloud applications. This architecture is ideal for creating a scene for the collection and preparation of large flows of information. Switching between topologies: useful for events in many ways (Khaddaj and Srour, 2016). It consists of four main segments: event lines, mediator, event channels, and event handlers. Agent topology: in this geography there is no central event between one event and another. There are two types of segments: representative processors and events.

2.4 **A review of Traditional Legacy (monolith) Architecture and cost maintenance, and modernisation**

Traditional or Legacy, often referred to as Monolith architecture are characterised by lack of support, are obsolete for current needs of a business or organisation, and difficult to maintain, improve, or integrate with new systems due to its architecture, underlying technology, or design. In a survey conducted among CIOs by Logicalis Global CIO, 43 percent views complex legacy technology a major barrier to digital transformation (Logicalis, 2017) and the answer is modernisation. Monolith architecture modernisation is the redesign and upgrade of old software system, and this normally results in maintenance costs becoming more and more expensive.

There has not been a formal definition for monolith architecture, but it is being compared with modular application architectureIt was recently that a better definition was given by (Dragoni et al., 2017) as "software application whose modules cannot be executed independently", rather the components are maintained and bundled together, distributed and deployed as a whole. The monolith as an architecture has some notable advantages, as it is easier to understand a specific process as all the elements are in the same codebase. On the other hand, the approach has some issues as well, for instance: any modification to any module of the application will require alterations in other modules as the modules are inter-linked (Dragoni et al., 2017). Creating more instances to Increasing the capacity of a software application to handle more requests in a monolith is difficult, as the entire system must be replicated instead of the individual module and more resource is wasted in the process (Ren et al., 2018). The size of monolith application grows year by year and as its complexity increases, its maintainability decreases, especially if good software design practices are not followed.

Currently, the cloud computing and hardware virtualisation are gaining relevance and provides an opportunity for competitive advantage, and it is becoming essential for companies to ensure flexibility on their systems. The need for organisational agility too is necessitating the need for "capacity to flexibly respond to changes in the environment by quickly adjusting products and service offerings" (Baskarada et al., 2018). Event-driven microservices architecture has been considered as a viable solution (Baskarada et al., 2018) as it combines scalability, maintainability, easy of deployment, less infrastructure cost, resilience, and reusability (Carrasco et al., 2018). In spite of its short comings, some authors are of the opinion that software development should start with a monolith, and then with time migrate it to event-driven microservices (Fowler, 2015). This pattern is called "monolith first". However, migrating a monolith can be a challenging process (Mishra et al, 2018). Monolith applications still drive most of the largest business sites worldwide, they have also become a bottleneck for innovation (Blockhuys, 2020).

*2.5 Benefits of Agile Adoption for Large Enterprises*

The rapidly evolving digital markets of today, as well as the increasingly intense struggle by organisations to attract customers, mean that they must evolve with speed and unprecedented agility (Gannon, Barga and Sundaresan, 2017). To cope with this reality, companies are looking for an advantage that will allow them to distinguish themselves from their competitors; and for many of them, this advantage consists in adopting agile methods.

Specialised studies, including Version One's State of Agile Survey, report fairly successful agile transformation experiences in companies. For less successful experiences, two main factors can be identified: the difficulty in changing the corporate culture and the difficulty in transforming middle management (Richter et al., 2017). Transforming culture is indeed not easy and can induce a strong resistance to change that must be managed with tact. As for management, it can quickly brace itself on old practices if the approach lacks listening and pedagogy.

The agile methods are piloting practice groups and project realisation. They originate from the Agile manifesto, written in 2001, which consecrates the term "agile" to reference multiple existing methods (Gannon, Barga and Sundaresan, 2017). Agile methods are more "pragmatic than traditional methods, involve the requester (client) as much as possible and allow great responsiveness to their requests. They are based on an iterative, incremental and adaptive development cycle and must respect four fundamental values broken down into twelve principles from which derive a base of practices, either common" or complementary.

Thus, to simplify, these methods consist in studying and quickly evaluating whether or not a solution is suitable for the organisation, in designing the solution, then in checking whether it is doomed to failure or, hopefully, to success, all this in the fastest and most economical way possible (Richter et al., 2017). Implementing this method for the design of microservices and their adoption on a company scale considerably reduces the risk as well as the time and resources invested, though the gains in terms of experience and learning are a factor of essential optimisation.

The culture and values conveyed by agility can conflict with those currently present in the company. It is believed that certain agile practices and values can prove to be in opposition to the culture of the company and have, by their nature, more difficulty in establishing themselves there than others (Kravari and Bassiliades, 2019). Agility is based on collaboration, a strong value that induces research into synergies, trust and teamwork. This can for example be found in contradiction with individual excellence, which induces a certain individual performance and the development of own expertise.

A company can decide to deploy agile practices on a large scale. If the managerial model is based on values of individual excellence, efficiency and a good dose of pragmatism, then certain agile practices will have some of the difficulties to be implemented. Indeed, agile managerial practices are based on autonomy, confidence and the development of team motivation (Li et al., 2018). Thus, the agile practices that convey these cultural values may prove to be out of step with the dominant culture of excellence of certain companies. During particularly tense moments inherent in the lifecycle of a project, managers have a key role in transforming an organisation towards agility. It is they who, by choosing to support and engage or not in this transformation, determine the outcome (Kravari and Bassiliades, 2019). As Jon Stahl explains in "Agile from top to bottom: leaders and leadership living agility", leaders must "embody the change they want to see born and set an example by reflecting agile values, by managing by example, by seeking to really understand the culture and finally, being as transparent as the teams they lead. This implies a posture of humility, empathy and confidence in the ability of the organisation to transform itself (Mayer and Weinreich, 2018).

The change must not come only from management, the teams are the best drivers of this change. Within a para-public organisation, they supported the deployment of a large number of agile practices and values. The transformation project was a success, but the biggest difficulty was the structuring of autonomous teams, insofar as the corporate culture was structurally based on a strong sense of hierarchy and control (Li et al., 2018). Managers had difficulty taking on the role of facilitator and letting go. It is a fairly frequent resistance to change that can have significant impacts on teams in terms of motivation and taking responsibility. And adopting agile practices in maintaining or creating a monolith architectural system has similar challenge.

Any agile methodology provides for the splitting of software development steps (Leppänen, Savaglio and Fortino, 2020), unlike the traditional method which foresees the total planning of the project even before its development, the Agile Manifesto advocates rather the setting of short-term objectives. Agile methods make it a point of honour to strengthen relationships between members of the project team, but also between the team and the client. It is for this reason that flexibility and flexibility in organisation are two fundamental pillars of agile methods.

## *2.6 Benefits of Implementation of Event-Driven Architecture for Large Enterprises*

The growing complexity of enterprise applications often leads to poor architecture, and the organisation spends more and more money on creating IT systems (Krämer, Frese and Kuijper, 2019). Event-driven architecture is designed to partially solve this problem by reducing the connectivity of software components or services. The purpose of event-driven architecture is to build systems in which components are loosely coupled.

The key concept in event-oriented architecture is the event. An event is a significant change in state. Events are passed between loosely coupled services and represent stages in some kind of business process. The service subscribes, observes events and responds to them. The Observer pattern helps to understand the concept of event-driven architecture (Al-Masri, 2018). In the Observer template, an object called a subject contains a list of objects called observers and notifies them of any change in state. Using events and observers allows you to make services loosely coupled.

Event-driven architecture helps create high-performance and highly accessible systems (Krämer, Frese and Kuijper, 2019). Designing systems asynchronous from start to finish allows you to minimise the time of thread blocking on IO operations and use the network bandwidth at full capacity. All observers can respond to event alerts in parallel, forcing multi-core processors and clusters to operate at the highest power. When a distributed system is running in a cluster, events can be delivered to any node in the cluster, providing transparent load-balancing and failover (Alwis et al., 2019).

The past few years have seen an increase in the popularity of microservice architecture (Al-Masri, 2018). There are many arguments against the use of microservices, but the most significant of them is that this type of architecture is fraught with uncertain complexity, the level of which depends on how you manage the relationship between your services and teams. Microservices assume the modularity of the code and the entire infrastructure (databases, etc.). In a correctly implemented microservice architecture, each service has its own infrastructure. Access to the user database (read and write) can be carried out only by the Users service (Boyer et al., 2018).

Microservices help solve organisational problems and give us the ability to manage the code base, which is changed by several teams (Mena et al., 2019). The separation of the code base prevents conflicts when making changes.

## *2.7 Recommendations for Best Practices and Adoption Coupling Compliance and Business Agility*

Large organisations as a whole will develop more slowly than their more flexible and advanced competitors and are increasingly immune to change (Boyer et al., 2018). The advanced and flexible structure of the Scaled Agile Framework (SAFe) finally becomes an integral asset that large associations can use to tackle problems that negatively affect corporate performance. SAFe offers large associations a system with which they can be more flexible so that their expectations are more accessible (Mena et al., 2019). Learn about the favourable circumstances and standards and how to best implement this system and its approaches.

SAFe contains many standards, procedures, and best practices that large associations can use to implement flexible strategies like Lean and Scrum to quickly create and transfer world-class resources and administrations (Strljic et al., 2019). SAFe is particularly well suited for complex tasks involving numerous large groups at the company, program and portfolio levels. The current module, SAFe 4.6, is based on a number of key functions that enable companies to effectively monitor and successfully adapt to disruptions in innovation to report instability, changes in customer needs and increasing success.

Agile and nimble leadership: Pioneers must stimulate and support significant change and work efficiency. Ultimately, it is the supervisory team that can influence individuals and groups to understand their hidden abilities (Strljic et al., 2019).

## *2.8 Bi-modal Operational Strategy*

The concept of a bimodal IT department has evolved ambiguously since the analytic company Gartner first introduced this model in 2014 (Xu et al., 2017). Some experts believe that this model, which implies parallel work on both everyday tasks and innovations, is just a practical recommendation for IT directors.

**2.8.1 Business Case for a Bimodal IT Model**

Using the bimodal IT model, if used correctly, can effectively convey to the rest of the business the real value of the IT solutions and help them get the most out of the technologies that are already available and used by the company (Christoforou, Odysseos and Andreou, 2019). If CIOs can continue to provide the main work of the department and at the same time create a business case for innovations, their ideas can find support from above (Christoforou, Odysseos and Andreou, 2019).

Direct business executives themselves are already able to use clouds and platforms like Amazon Web Services to launch new services on demand. Enterprises of traditional specialisation, such as banks and insurance companies, establish units in their activities that are dedicated to digital business forms. These organisations have realised that they must compete effectively (Reilly, 2020).

Such distinguished structures are a good example of a bimodal IT model: the firm continues to engage in core business, but at the same time tries to develop something new (Leppänen, Savaglio and Fortino, 2020). However, it is imperative that CIOs do not see the bimodal IT model as a lifesaver. In the best case, this model should be used as a transitional stage, while the leadership in the meantime should prepare for the continuous implementation of change (Avritzer et al., 2020).

There is some element of danger in the bimodal IT model. CIOs of traditional architecture views who are not keeping pace with the requirements of globalisation can come to a meeting of the board of directors, talk for a long time about the need to support basic IT operations and thus justify the absence of changes (Ren et al., 2017). IT infrastructure existed as part of the enterprise only because we could not instruct anyone on the side to deal with it. Now we have the clouds, and this has become possible.

The business is undergoing fundamental changes (Leppänen, Savaglio and Fortino, 2020). In the financial sector, a competitor-novice can come up with a completely new type of product, and in the traditional architecture business model, it can take a whole year to respond adequately. So, organisations have to get used to the changes, take them for granted and work with them in mind (Oparin, Bogdanova and Pashinin, 2019).

Thus, a bimodal approach to information technology can help IT leaders convince others of the powerful potential for innovation. However, according to Hudson as stated earlier, CIOs should not really rely on a bimodal IT concept (Ren et al., 2017). The managers of business units have already been overloaded with overly promoted technical terms, and the vague wording can hardly improve the negative impression of everything related to information technology.

It is necessary to bring the project to the market carefully, taking into account all the nuances. They need to interest people and give them the opportunity to take a fresh look at information technology (Oparin, Bogdanova and Pashinin, 2019). The task of CIOs is to help all business participants realise that change through technology is great, like migrating from a monolith to microservices event-driven architecture. A bimodal model can give IT projects a more positive image.

The key step to success is to achieve positive results using a bimodal IT model. In the end, businessmen need exactly the results. Using a bimodal approach, you can lower the barrier to maximise the benefits of IT solutions and, by experimenting, achieve the positive results that business needs so much (Jackson and Lui, 2017). A bimodal IT model simply acts as a tool. With a competent approach to business, and they need to figure out what this tool is and how it can be used to improve the business.

## 2.9 A Review of Market adoption of Event-driven Architectures

As a result of the obvious challenges with monolith architecture, and the critical need to maintain competitive advantage, companies are migrating to microservices event-driven architecture. Notable of them are LinkedIn, Amazon, Netflix, IBM, Uber, SoundCloud and others (Ren et al., 2018), and a brief review will be done on Netflix. A microservice architecture consists of autonomous components that can be deployed in separation. "Microservice event-driven architecture is a distributed application where all its modules are microservices" (Dragoni et al., 2017). It consists of a distributed application in which its behaviour depends on the communication, composition, and coordination of its components via messages (Dragini et al., 2017).

Microservices is one of the latest movements in software architecture, particularly the event-driven frameworks for creating loosely coupled components. It is an evolution of the oldest concept of Service Oriented Architecture (SOA) which relies on middleware like Enterprise Service Bus (ESB), and Web Service Description Language (WSDL), microservices on the other hand rely only on simple technologies like Representational State Transfer (REST). Furthermore, SOA is normally viewed as an integration solution of already existing systems, while microservices are typically used to develop new and individual software systems (Jamshidi et al., 2018).

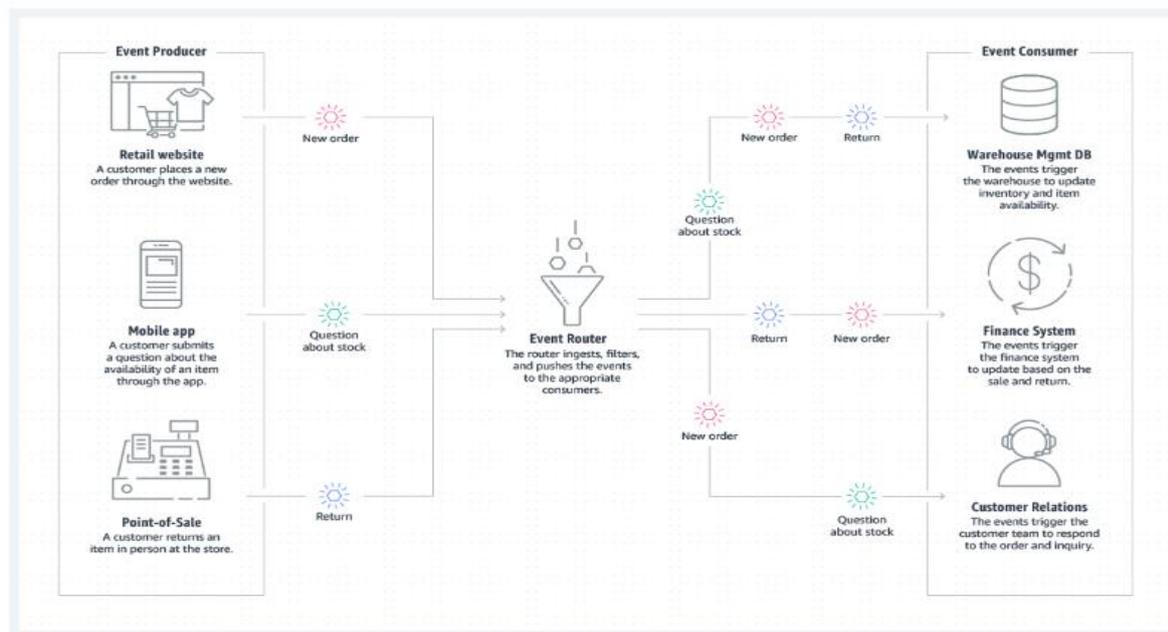

Figure 2.3: An event-driven architecture for an e-commerce site (AWS, 2020)

2.10 **Case Study of Netflix**

Netflix is one of the big enterprises that has migrated to microservices from the traditional monolith application in 2009. Today, Netflix has more than 800 microservices running, and serving around 2 billion requests daily, with about 20 billion internal API hits in these microservices.

When Netflix started moving from a monolith to microservices architecture, the term "microservices" did not exist. They used to have a big monolith application and the movement to microservices started when the application experienced a major database bottleneck that shut off operations for three days in which they were unable to ship DVDs to their clients. This was because the databases in monolith systems have single point of failure. The situation inspired them to consider migrating from 'vertically scaled single point of failure' (Meshenberg, 2016), referring to the relational databases in their datacentre containing monolith systems, towards highly reliable, horizontally scalable, distributed systems in the cloud. They picked Amazon

Web Services (AWS) as their cloud provider. AWS provided them with the greatest scale and the broadest set of services and features.

They started with moving the movie encoding code base, a non-customer-oriented application, to small new microservices. This is the best method to ensure minimal or no disruption to the online services. (Sharma, 2017) suggests to always move non-customer-oriented components from monolith to microservices architecture first. This is because, when such components fail, there will be less impact on traffic and the customer experience on the production. After successfully moving non-customer-oriented components, they started moving customer-oriented components. Netflix completed migrating their application to the microservices architecture in 2016 and shut off their remaining data centres.

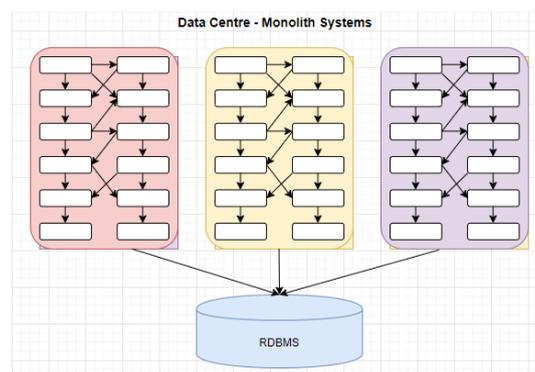
Figure 2.4: Netflix monolith architecture

The diagram above depicts a monolith architecture, similar with Netflix's system back in 2008 that crashed and shut off business for four days. The system has a single point of failure and once there is a problem, the whole system crashes.

 2.11 Issues Observed and Recommendations for Best Practices

Some development teams find advantages over monolithic in microservice architecture (Reilly, 2020). Others consider microservices an unnecessary load that reduces performance. Like every architectural style, microservices have both advantages and disadvantages. This study presents the specific cases where it is better to use a microservice architecture so that the choice is reasonable (Chen, Li and Li, 2018).

The first advantage of microservices is the loosely coupling of the modules. This is an important, but rather strange advantage, since in theory it is not clear why microservices should have more flexible module boundaries than monolithic code (Fowler, 2015). A good modular structure is useful for each program, but only with the growth of the development team and the increase in the size of the software does it become vital. Proponents of microservices immediately recall Conway Law, according to which the system reflects the communication structure of the organisation into which it was developed.

With the growth of teams (especially if they are geographically distributed), it is important to structure the software so that communications between individual groups are less frequent and more formal than within groups. And microservices allow you to comply with this pattern of communications, providing each team with relatively independent elements of the system.

As stated above, there is no reason to believe that a monolithic system cannot have a good modular structure (Guiotto et al., 2015). But we know that this happens extremely rarely, and the most common architectural concept for monolithic systems is the "Big Ball of Mud", which is increasingly becoming the reason for turning towards microservices.

Separation in this case works because the boundaries of the module provide a barrier to inter-module links (Masala et al., 2017). In a monolithic system, as a rule, such a barrier is easy to circumvent. This allows you to quickly add new features, but with widespread use, this approach undermines the modular structure and reduces team productivity. Separating modules into separate services makes the boundaries more stringent and getting around them becomes more difficult.

In addition, early adopters of new ideas are usually more talented, which gives an additional delay for evaluating the modular benefits of microservice systems written by medium-sized development teams. Everything that can be said now is based on the testimonies of development teams who have used this style. They say that supporting modules with microservices is much easier for them.

At that time, "microservice architecture became useful, since the system was able to cope with the rapid influx of developers, and the development team" was able to use a large number of teams much easier than with a monolithic approach (Rezaei et al., 2017). As a result, the project accelerated to a greater extent than would have been expected with a monolithic approach, which allowed the team to catch up with the schedule. The overall result, however, was negative at that time, because software took more man-hours than it would be on a monolithic architecture.

Microservices are prone to inconsistency due to their respectable "insistence on decentralised data management". In a monolithic system, you can update many objects in one transaction. With microservices, it will take several resources to upgrade and distributed transactions. As a result, now developers must remember the possibility of inconsistency and think about how to determine the moments of desynchronization before embarking on programming task.

The monolithic world is not free from these problems. As the system grows, the need to use caching to improve performance grows, and invalidating the cache is another difficult task (Pitakrat et al.,2018). Most applications require offline locks to avoid long-running database transactions. External systems require updates that cannot be negotiated using the transaction manager, and business processes are often more tolerant of inconsistencies than could be assumed, because companies usually prefer the availability of consistency.

The use of microservices, amongst other things, was caused by the complexity of deploying large monolithic systems, where a small change in part of the system can lead to failure of the entire deployment (Ge et al., 2017). There are also many examples of unsuccessful attempts to use microservice architectures in independent deployments, when releases of several services need to be carefully coordinated.

In the literature review, there was no conclusive or satisfactory answer found for the performance differences and the cost of maintenance of the two architectures that could help us determine the scope and impact of migrating to microservices event-driven architecture. Therefore, this study aims to study this which is the focus of this research work.

# Chapter 3

# Methodology

This research work is conducted by reviewing existing literature from peer reviewed articles, journals, documents published by major middleware companies, and books authored by software professionals and researchers on the relevant subject matter. The answers to research questions (RQ1and RQ2) are sought using empirical study of relevant literature from previous studies and from publications by companies, including developing prototypes of the two architectures to be used for an experiment to obtain their performance metrics, as well as obtaining and analysing cost of maintenance to enable the comparison of what an organisation likely spends maintaining a monolith or a microservice application per month (Kramer, Frese and Kuijper, 2019).

For the methodology, this work chose to use an experiment to obtain numerical data of the performance metrics, which could not be obtained from surveys, case studies and interviews. To conduct the experiment, a monolith and microservice application prototypes were developed and integrated with Dropwizard metrics, an open source library for measuring the performance of web applications and paired with SpringBoot actuator to enable measurement. An Apache JMeter was also installed and used to simulate number of users sending requests to the applications simultaneously, and the response time taken. Performance aspects such as RAM usage, CPU usage, Response Time, Scalability and Latency were considered. The prototypes developed were basic in nature and hosted locally and the sole purpose of creating the prototypes is to enable simulation and collect required data for the analysis. The monolith application was hosted on localhost:8080, while the microservice application consisting of a gateway frontend and two microservices were hosted on a Docker container.

3.1 **Experiment Design**

This begins with the Experiment definition, explaining the purpose of conducting the experiment, sketch description of the architectures and setting. The Independent and Dependent variables are then presented, followed by experiment steps and what tools were used to gather the performance data.

3.1.1 **Experiment Definition**

In the literature review, (Chen, Li and Li, 2018) discussed about situations where it is better to migrate to microservices and other cases where it is best to use monolith architecture. (Pitakrat et al., 2018) also pointed out an issue regarding the behaviour of monolith systems, observing that as the monolith grows over time and the number of request calls increases, its performance to drops and the system becomes increasingly slower. This study aims to analyse this using system metrics and conducting an experiment seemed to be the best way to obtain actual performance differences.

3.1.2 **Research Setting**

The research setting simulates two applications: a monolith and a microservice, that have user interface, REST API, business logic and database, and integrated with a Dropwizard Metrics paired with SpringBoot actuator, an open source library for measuring the performance of web applications. The applications were developed using angular client-side code for frontend, java application to integrate with the backend.

3.1.2.1 **The Monolith Application**

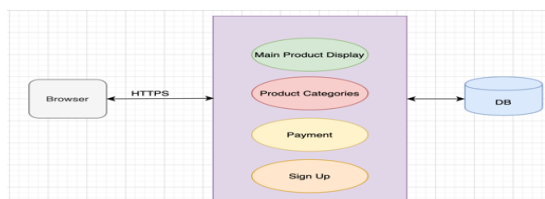

Figure 3.1: Monolith architecture running on localhost

The monolith application was built using angular, html and CSS for user interface, Java application to integrate with the backend MySQL database, integrated with DropWizard Mtrics and SpringBoot Actuator to enable measurement of system metrics. The frontend has ecommerce entities comprising products, product category, customer, product order, order item, invoice and shipment. However, it is important to note that the invoice in monoliths has performance issue when the number of user request calls is high, thus slowing down the system.

3.1.2.2 **The Microservice Application**

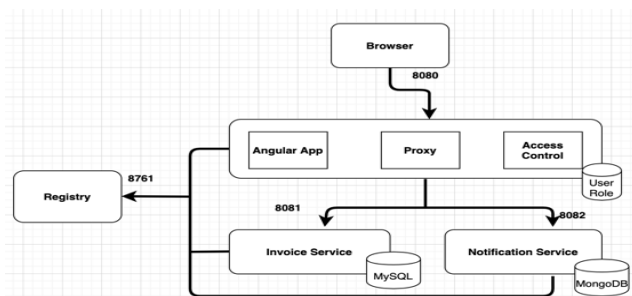

Figure 3.2: Microservice architecture running on Docker machine

Unlike the monolith application, the microservice application has application Gateway which is the frontend developed using angular, html and CSS with MySQL database to store user roles and just like the monolith, it has entities of products, product category, customer, product order, order item and shipment. However, to address the performance issue identified in the monolith application with the invoice entity, it has been isolated into a stand-alone microservice and integrated with a MySQL database, in addition to creating a notifications microservice and integrated with a mongo database. The registry is a separate application in the Docker machine that links the application Gateway and the independent microservices. A DropDown Wizard metrics and SpringBoot Actuator was integrated to enable reading of performance of the application.

### 3.1.3 Independent Variables

The independent variables for the experiment is the number of user requests simulated using JMeter while the applications were running.

### 3.1.4 Dependent Variables

The dependent variables for the experiment are the average response time and line response time measured in milliseconds.

### 3.1.5 Experimental Steps

The experiment steps consisted of development of monolith and microservices applications, then measuring test cases and collecting performance data.

### 3.1.6 Measurement Instrument

To measure the user response time, the latest versions of JMeter and dropwizard metrics were used.

### 3.1.7 Measurement Objects

The measurements were conducted in a test plan for JMeter and Dropdown wizard metrics. For the JMeter, a ramp up period of 60seconds were used to run each test case for two minutes, followed with a delay of two minutes before starting the next. The JMeter represented users making a REST request to the user interface in an ecommerce store. This request could be searching a product, placing an order, and transacting an invoice. JMeter is used to distribute load across systems and analyse performance, such as response time and error rate.

Two test cases s1 and s2 were performed for the two architectures:

S1 – 10 simultaneous user requests per minute

S2 – 1,100 simultaneous user requests per minute

### 3.1.8 Data Collection

After starting and running the monolith application in localhost 8080, and the microservices in Docker container respectively, the test plans were conducted and read, and captured snapshots of performance metrics, which was then extracted into Excel for analysis.

3.1.9 **Data Analysis**

The data was extracted into Excel where it is analysed and summarised into tables and diagrams.

The dropdown wizard metrics tool provided the system metrics such as RAM, CPU usage, while the Apache JMeter provided performance aspects like average response time, 90%-line response time, error rate and throughput. The purpose of analysing these performance metrics is to help in understanding which architecture performs best when number of user request calls increases. Only metrics that will help in answering the research questions are used.

The diagrams and tabulated results are used to help in finding the differences between the two architectures. The result from the experiment is presented in Results and Analysis.

3.1.10 **Validity Threats**

The first validity threat is that the experiment used a small number of users for the sample test, and this is because the experiment was conducted in a single computer rather than a distributed system. However, with more users, a higher scalability difference could have been observed. The reason of limiting the sample test case to 1,100 simultaneous user requests is that the system could not handle more than that at a go. This limitation could be handled using a distributed system.

Secondly, the applications were run locally on localhost and docker. The results might vary if hosted on the cloud servers.

# Chapter 4

# Results

**RQ 2**: What are the scope and impact of migrating to microservices event-driven architecture, including using agile frameworks?

To answer this research question, experimental data comparing performance characteristics is presented, as well as data supporting cost of maintenance and impact on agile teams obtained from industry. Firstly, the experimental results will be presented first, then the data comparing the cost of maintenance and agile teams will be presented next.

4.0 **Experimental Results**

4.1 **Response Time**

Response time was measured for the sample cases s1 and s2 on the monolith and microservice systems. Response time is the elapsed time prior to sending the request, and immediately after the response is received. To measure response time, test plans were created to run a constant workload of 10 and 1100 requests per minute for a duration of 5 minutes on each architecture, resulting in a total of 55 and 5500 requests respectively. This is within the tolerant level of the systems without crashing. For response time, two samples s1 of 10 requests and s2 of 1100 requests were made for both architectures. When 10 user requests were made, the response time for monolith architecture was 283.7 milliseconds, far lower compared to microservices that took 322.9 milliseconds. Thus, when the number of user requests is lower, monolith performs better than microservices. On the other hand, a request call of 1100 showed a contrasting result. The response time for monolith was 285.33 compared to microservice which returned 210.0, showing a decrease in 75.33 milliseconds. This proves why as monolith systems grow over the years their performance decreases and becomes a bottleneck to maintain.

Microservices on the other hand are robust and can handle increasing request calls. The table and graph show the test results.

| Sample | Monolith (milliseconds) | Microservice (milliseconds) |
|---|---|---|
| S1 (10 requests) | 283.7 | 322.9 |
| S2 (1100 requests) | 285.33 | 210.0 |

Table 4.1: Response time for 10 and 1100 request calls

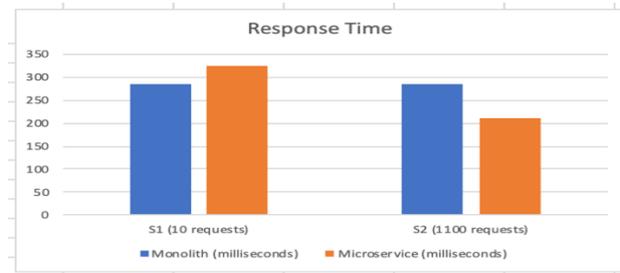

Figure 4.1: Response time graph

### 4.2 Line Response Time

90% Line response time is the range above which 90% of request calls are faster. Just like in the response time discussed above, the monolith architecture performs best when the number of request calls is fewer, and worse when the request calls increases. Microservices on the other hand does not show optimum advantage with fewer request calls. However, its robustness and scalability is noticed when number of request calls grows.

| Sample | Monolith 90% (ms) | Microservice 90% (ms) |
|---|---|---|
| S1 (10 requests) | 308 | 403 |
| S2 (1100 requests) | 354.33 | 275.33 |

Table 4.2: Line Response time for monolith and microservices

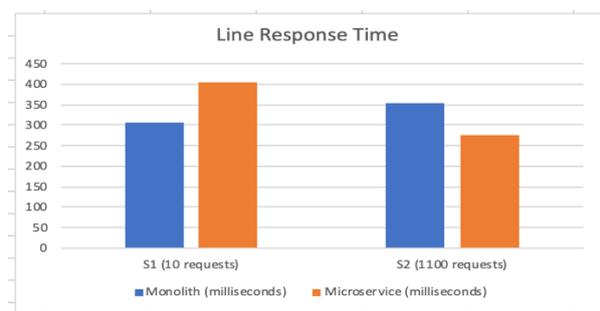

Figure 4.2: Line response time graph

4.3 **Error Rate**

Error rate was measured for the sample cases, s1 and s2. Error rate depicts the number of failed requests that returned a response with status code 500, implying an internal server error. Test plans were created to run a constant workload of 2000 requests per minute for a duration of 5 minutes on the monolith and microservice, resulting in a total of 10000 requests. This was done purposely to stress the systems to the point of crashing for the purpose of analysing how the two different architectures handle errors. To measure error rate, the use case for monolith was represented with s1 and use case for microservice s2. The test performed for the monolith, s1, returned an error rate of 62%, while the microservice, s2, returned 17% error rate. This means that of the 10000 requests sent, the monolith architecture has a 38% successful response with a status code of 200, resulting in 3800 successful requests. The microservices on the other hand, has 83% response with a status code of 200, that is about 8300 successful requests. The results are presented in the table and graphically below.

| Use case | Sample size | Error Rate (%) |
|---|---|---|
| S1 | 10000 | 62% |
| S2 | 10000 | 17% |

Table 4.3: Data showing error rate for monolith (s1) and microservice (s2) samples

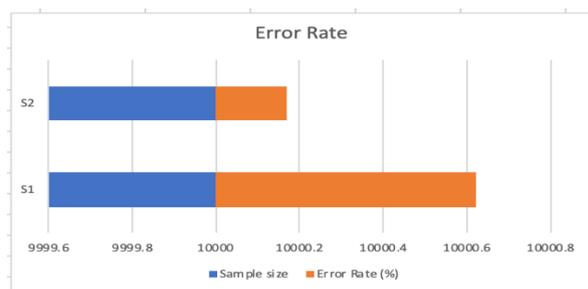

Figure 4.3: Error rate graph.

### 4.4.2 Comparison of Cost of Maintenance
#### 4.4.2.1 Infrastructure Costs of Monolith Architecture on AWS

| Service | Cost per hour (USD) | Quantity per Month | Cost per Month (USD) |
|---|---|---|---|
| Web application. EC2 Instance c.4 large | 0.110 | 720*4 | 316.80 |
| Web application. RDS Instance db.m4.large with Single AZ | 0.095 | 720*1 | 68.40 |
| Web application. ELB Instance | 0.025 | 720*1 | 18.00 |
| **Monthly infrastructure costs** | | | **$403.20** |

Table 4.4: Monolith infrastructure cost

Table 4.4 above gives the cost per hour, quantity per month and cost per month for the monolith architecture in AWS which shows that the cost per hour of monolith architecture is 0.110 USD and for web application - Relational Database System (RDS) instances it is 0.095 and for web application - Elastic Load Balancer (ELB) the cost is 0.025 per hour. Similarly, after counting the price per month, it goes to a total amount of 403.20 US dollars for a monolithic architecture on Amazon web services.

#### 4.4.2.2 Infrastructure Costs of the Microservice Architecture on AWS

| Service | Cost per hour (USD) | Quantity per Month | Cost per Month (USD) |
|---|---|---|---|
| Microservice S1. EC2 Instance c4.large | 0.1101 | 720*3 | 237.816 |
| Microservice S1. ELB Instance | 0.0225 | 720*1 | 16.20 |
| Microservice S2. EC2 Instance t2.small | 0.0159 | 720*1 | 11.448 |
| Microservice S2. RDS Instance db.m4.large with Single AZ | 0.095 | 720*1 | 68.40 |
| Gateway. EC2 Instance m4.large | 0.067 | 720*1 | 48.24 |
| **Monthly Infrastructure Costs** | | | **$390.96** |

Table 4.5: Microservices infrastructure cost

Similarly, table 4.5 gives the price of hourly and monthly bases of microservices architecture which is 0.1101 for microservices system 1 Elastic Compute (S1 EC2) per hour and for microservices system 1 Elastic Load Balancer (S1 ELB) per hour is 0.0225. Also, for microservices system 2 (S2 EC2) instance the price goes to 0.0159. Summing up all the cost for monthly services of micro service architecture from Amazon Web services, the total cost that the user has to afford comes to be 390.96 US dollars.

### 4.4.3　Communication Support for Agile Frameworks

A unique advantage of microservices over monolith is the size of the development team. The number of communication channels increases with the number of team members. The formula for the number of communication channels is n(n-1)/2.

For example, a team with 20 developers will have 190 possible channels of communication.

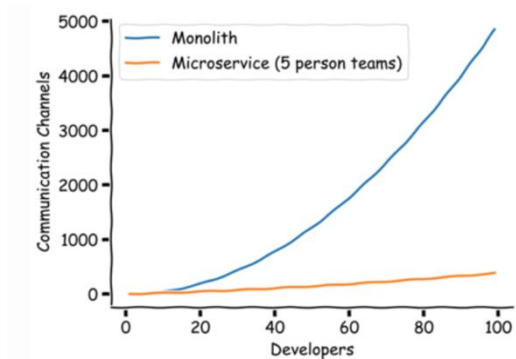

Figure 4.4: Monolith vs. Microservices communication channels as number of developers grows

As the graph shows, from size of 10 developers, the microservice model shows a unique advantage over the monolith.

# Chapter 5

# Analysis and Discussions

This research work discussed the challenges with monolith architecture and presented reasons organisations are migrating to microservices. However, many other organisations are reluctant to migrate to microservices, and that is where this research work comes in: to answer the questions why migrating to microservice is beneficial to the modern technology. The study objectives and research question one (**RQ 1**) were met in the literature review, while the experiment, cost of maintenance and agile frameworks were met in research question two (**RQ 2**). In this chapter, an attempt will be made to analyse how the research questions were addressed.

**RQ 1**: *How to migrate to microservices event-driven architecture while not disrupting ongoing business operations that depend on the traditional architecture system?*

5.1 **Migrating to Microservices with Minimal or No Disruption to Business Operations**

Deducing answers from the literature review and from the prototypes developed for the experiment, the first method to migrate from monolith to microservice is following what (Strljic et al., 2019) observed and following the recommended solution. According to them, the biggest problem in changing monolith to microservice lies in changing the communication template. The recommended solution is to reduce the number of communications between modules. This relates with the experiment carried out in this project in which the invoice module from the monolith was separated and converted to a separate microservice, in addition to creating a notification microservice. This reduced the number of communication channels.

Another method an organization can use to migrate to microservices with minimal or no disruption to ongoing business functions is to adopt the bimodal IT strategy. The strategy encourages parallel work on systems. This means continuing to maintain the monolith system while building the proposed microservice at the same time (Leppänen, Savaglio and Fortino, 2020). This method is obviously expensive as it implies creating a new system from scratch.

The third method is following the Netflix case study. Netflix migrated their monolith to microservices moving the movie encoding code base, a. non-customer-oriented application, which ensured minimal or no disruption to the online services. (Sharma, 2017) recommends moving the non-customer-oriented applications first, noting that even when such components fail, there will be less impact on traffic and the customer experience on the system. After successfully migrating the non-customer-oriented components, start migrating customer-oriented components.

The above three methods are the findings that attempted answering the research question 1 (RQ 1).

**RQ 2**: *What are the scope and impact of migrating to microservices event-driven architecture, including using agile frameworks?*

To answer this research question, the experimental results, cost of maintenance and agile communications will be analysed.

5.2 **Response Time**

At 10 request calls, the monolith system had a response time of 283.7 milliseconds compared to microservices which took 322 milliseconds. Thus, the microservice had a slower response time of 38.3 milliseconds. On the other hand, at 1100 request calls, the monolith system had a response time of 285.33 milliseconds versus microservices response time of 210 milliseconds. This time around, the microservice responded faster with about 75.33 milliseconds.

For Line Response Time, there was a related trend: the monolith system under 10 request calls responded within 308 milliseconds, while microservices responded within 403 milliseconds.

That is, monolith responded faster, and microservices was slower with 95 milliseconds. As the request calls was increased to 1100, however, monolith responded within 354.33 milliseconds, while microservices responded within 275.33 milliseconds. That is, the microservice was faster with about 79 milliseconds.

The trends show that when the number of users requesting calls to the system are few, such as 10 request calls, the monolith system will always outperform microservices. On the other hand, as the users visit to the system increases, sending increased request calls, such as 1100 calls, monolith system will show decreased response time, while the microservice will be able to handle such requests faster.

This finding shows that if a proposed system, like a web application is not intended for large users sending in requests at the same time, and there is no likelihood that the system would grow in future, the it is recommended to use a monolith architecture. On the other hand, if the proposed application is an e-commerce with projected large users visiting and placing requests at the same time, for example Black Friday deals, then a microservice architecture system is highly recommended. In summary, the bigger the system, the more advantageous microservices architecture will have over the monolith architecture.

5.3 **Error Rate**

The results show a monolith system, s1, with an error rate of 62% compared to a microservices system, s2, with error rate of 17% during the crash experiment. The microservice, s2, has a significant higher fault tolerance than the monolith system, s1. This finding has supported the theory that monolith systems have a single point of failure as all the modules are interlinked and connected to a single database (Ge et al., 2017), hence a fault in a module may cause the whole system to crash. Microservices on the other hand, are loosely coupled and each microservice has own database and as a result, when there is an issue with a single microservice or database, it rarely affects the whole system. The point of failure in a microservice can be easily isolated and reconfigured.

5.4 **Cost of Maintenance**

As the cost of maintenance shows in table 4.4 and table 4.5, for money spent on resources, microservices deliver more throughput and more precise scaling. For example, if a monolith is using all the resources, to handle more connections means bringing in another instance. On the other hand, if a single microservice uses all the resources, only this service will need more instances. Microservices are less resource intensive, and this saves resources.

From the tables, the monthly cost of microservices architecture was $390.96 versus the $403.20 for the monolith architecture. The result shows that the microservice architecture has a reduced cost of $12.24, and when using AWS Lambda, the costs can be reduced further. For example, when a monolith is running on the biggest AWS instance, m5.24xlarge instance with 96 CPUs and 384 GB RAM, the present cost is $2,119.19 per month for the instance running 24 hours 7 days (24/7). For the same money, 12 c5.2xlarge instances with 8 virtual CPUs and 16GB of RAM each running 24 hours 7 days (27/7) can be bought to run microservices (Kainz, 2020).

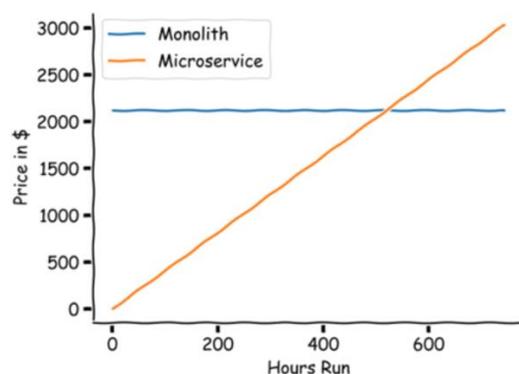

Figure 5.1: Price of 12 on-demand microservice containers vs. a dedicated large monolith container (Kainz, 2020)

5.5 **Benefits of Agile Methods**

Agile methods and frameworks give organisations competitive advantage, improves communication and collaboration among team members and makes it easier to adopt modern practices in the industry (Gannan, Barga and Sundaresan, 2017). As the graph in figure 4.4 shows, communication channels are increasing as team size goes up in monoliths but increases relatively slower in microservices. For instance, as the graph shows: in monoliths, a team of 20 developers have a communication channel of 190; a team of 40 developers have a communication channel of 780, a team of 60developers have a communication channel of 1770, and so on. On the other hand, in microservices, a team of 20 developers have a communication channel of 50; a team of 40 developers have a communication channel of 200, and a team of 60 developers have a communication channel of 400 and so on which are quite significantly lower in comparison to the teams in monoliths. Microservices with loosely coupled and isolated microservices enables the breaking of big teams into many smaller teams, thereby significantly reducing the communication channels. As a result, microservices can cope with rapid influx of developers (Rezaei et al., 2017), and large development teams can collaborate easily. Monoliths on the other hand are more expensive to maintain considering the challenge in coordinating communication and collaboration among increasing teams.

# Chapter 6

# Project Management

**6.1 Project Schedule**

The project schedule was defined during the project proposal as an approximate timeline to be followed, but with options for alteration as possible, because the researcher did not have a clear structure and definite direction for the project at outset. The table below shows the original project timeline:

6.1.1 **Project Timeline**

|   | **Activity** | **Duration** |
|---|---|---|
| 1 | Project topic definition | 26$^{th}$ May-18$^{th}$ June |
| 2 | Secondary data collection and processing | 27$^{th}$ May-25$^{th}$ June |
| 3 | Primary Data cleaning, processing and storage | 26$^{th}$ June-5$^{th}$ July |
| 4 | Discussion of findings | 6$^{th}$ July-9$^{th}$ July |
| 5 | Working on CW2 | 9$^{th}$ July-16$^{th}$ July |
| 6 | Documentation | 14$^{th}$ July-10$^{th}$ August |

Table 6.1: Timeline

Following feedback from the research proposal, the project topic had to be narrowed further, and this means that project objectives, research questions and methodology had to be altered as well.

Further, the initial research methodology was to obtain big data on the project topic from dataset sites. Unfortunately, the researcher could not find available dataset suitable for the research. As such, project supervisor advised creating prototype applications and generate required data. Therefore, the research methodology was changed from data collection and processing to creating prototypes and performing experiments. This impacted the original project plan and caused activities to be behind schedule.

6.1.2 **Current Project Plan**

|    | **Activity**                                                 | **Duration**                                |
|----|--------------------------------------------------------------|---------------------------------------------|
| 1  | Final project topic                                          | 19$^{th}$ June, 2020                        |
| 2  | Literature search continues                                  | 27$^{th}$ May-30$^{th}$ June                |
| 3  | Creating prototypes for monolith and microservice applications | 9$^{th}$ July-15$^{th}$ July              |
| 4  | Experimentation and data collection                          | 15$^{th}$ July-16$^{th}$ July               |
| 5  | Submission of CW2                                            | 17$^{th}$ July 2020                         |
| 6  | Presentation vis Teams                                       | 22$^{nd}$ July 2020                         |
| 7  | Data Analysis and discussion                                 | 23$^{rd}$ July–5$^{th}$ August              |
| 8  | Solution, conclusion and recommendation                      | 6$^{th}$ -8$^{th}$ August                   |
| 9  | Final Report                                                 | 9$^{th}$ – 13 August                        |
| 10 | Final submission                                             | 14$^{th}$ August 2020                       |

Table 6.2: Current timeline

## 6.2 Work Breakdown Structure

For the work breakdown structure, the researcher made use of sprint board in Trello (trello.com) where the tasks are outlined in order of Backlog, Sprint Backlog (planning), To Do, In Progress and

### 6.2.1 **Task – complete**

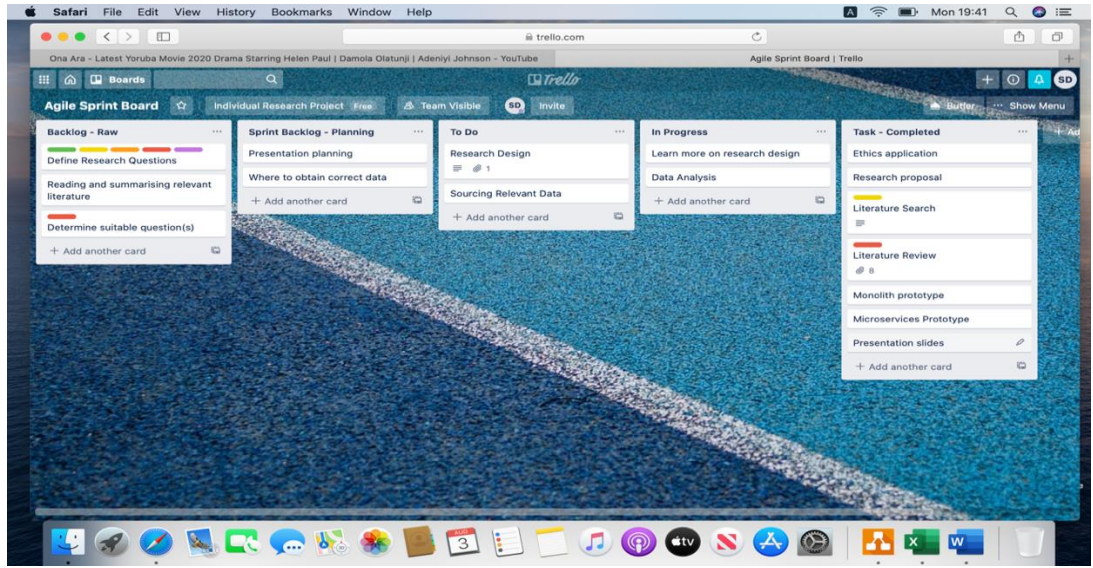

Figure 6.1: Work breakdown structure using sprint board

### 6.2.2 **Gantt Chart**

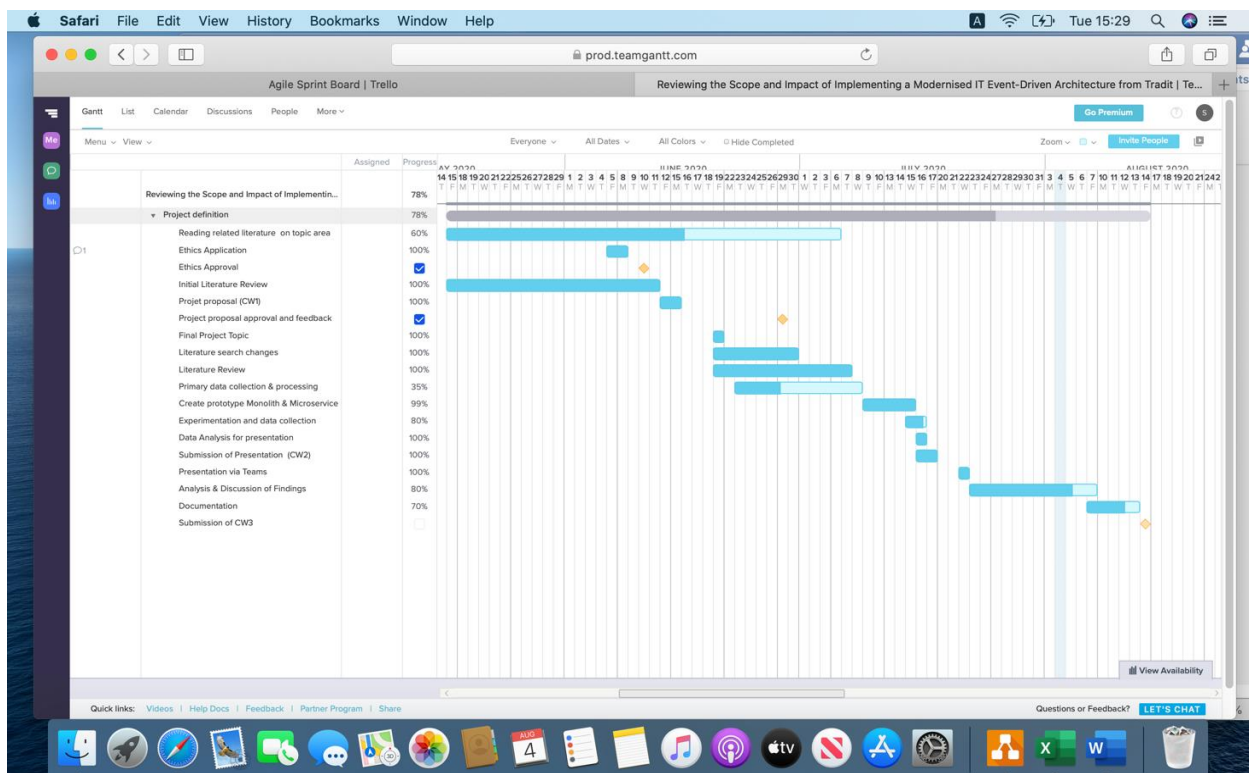

Figure 6.2: Gantt Chart showing project steps

6.2.3 **The changes in project plan includes the following**:

i. Change in project topic. The initial project topic was too broad and had to be narrowed down.

ii. Change in the literature search in order to suit the new topic

iii. Change from data collection to experimentation as the means to obtain the required primary data for analysis and discussion

6.3 **Risk Management**

The university of Oregon, uoregon (nd) has listed potential risks students undertaking research are likely to face. These include:

i. Physical risks: This could manifest in physical discomfort, pain, injury, or illness that may result from the methods and procedures of the research.

ii. Psychological risks: This may include the effects of anxiety, depression, guilt, shock and loss of self-esteem. Others are sensory deprivation, sleeplessness, use of hypnosis, deception or mental stress.

iii. Social or Economic risks: this may include alterations in relationships with others, financial issues and loss of employment.

iv. Loss of Confidentiality: This is a risk associated with researches involving human subjects where confidentiality of presumed information is expected to be maintained judiciously. Great care needs to be taken to avoid inversion of privacy and personal dignity.

v. Legal risks: This exists when the research methods are in a way that the subject will be liable for a violation of the law.

## 6.4 **Risk Assessment**

The researcher used Risk Assessment Tool available at myriskassessment.co.uk to define and analyse possible risks, as well as possible precautions and actions to be taken if the risk happens.

### 6.4.1 **Physical Risk**

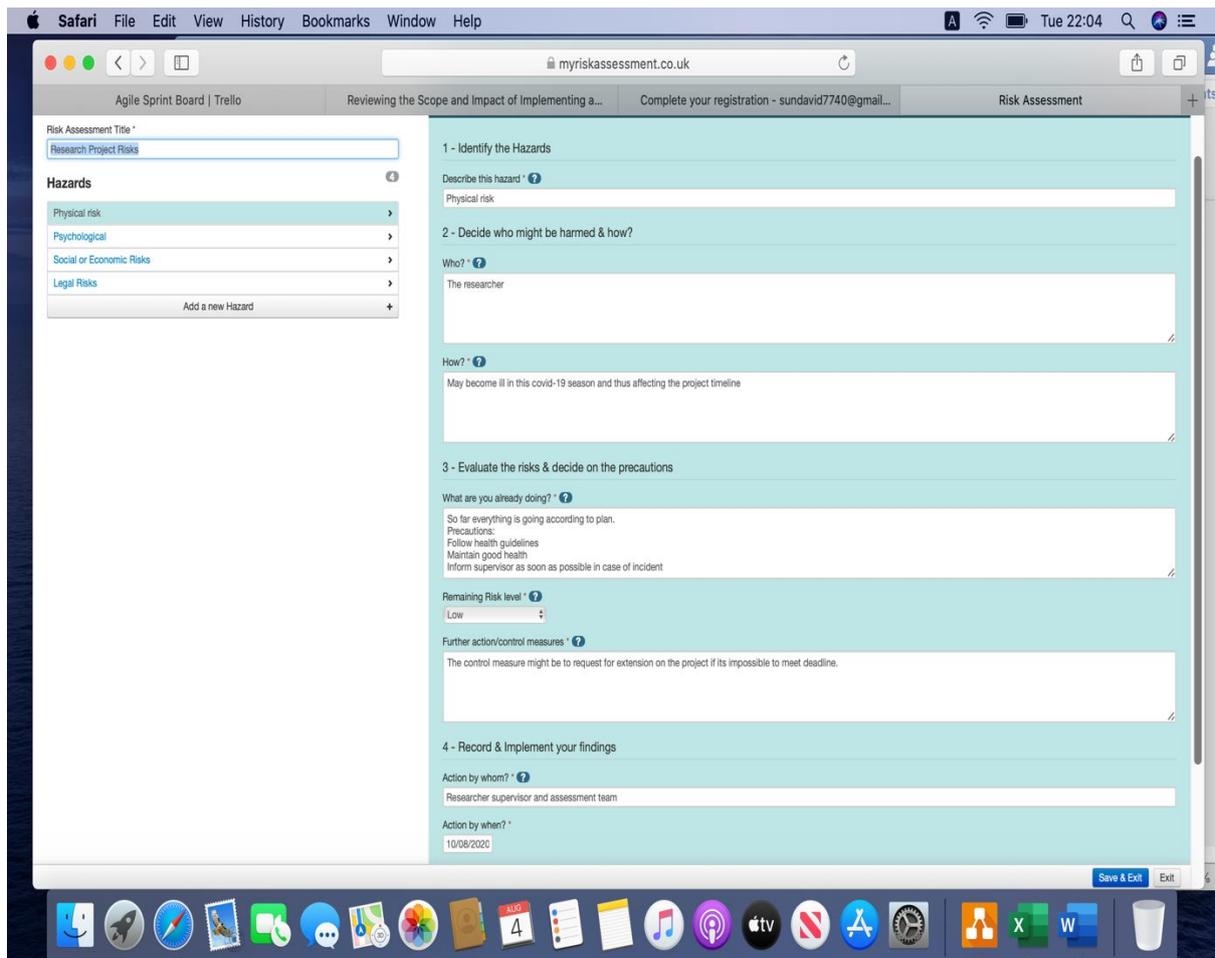

Figure 6.3: Physical risk

### 6.4.2 **Psychological Risk**

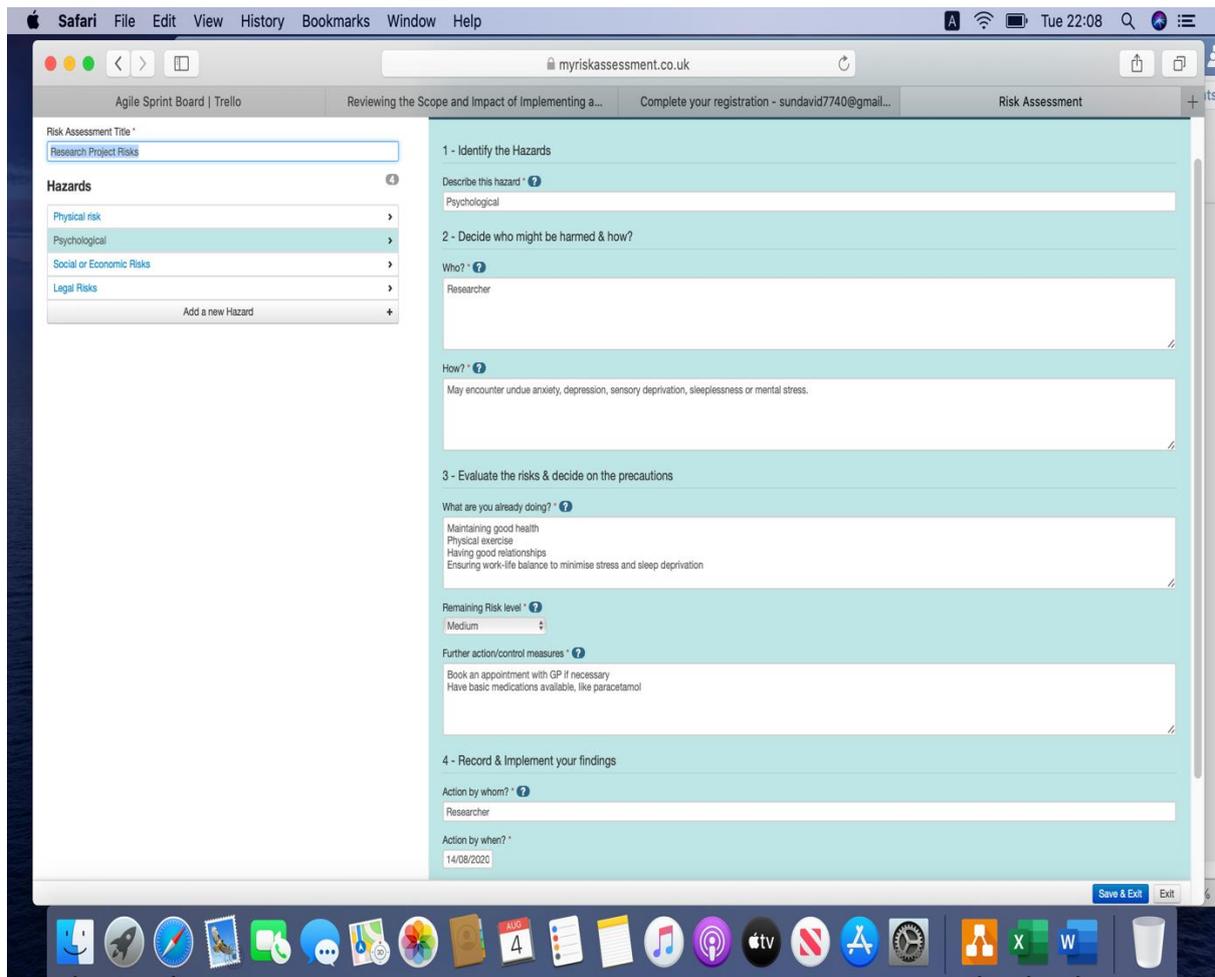

Figure 6.4: Psychological risk

6.4.3 **Legal Risk**

Figure 6.5: Legal risk

So far, the researcher has not experienced any serious risks, except the unexpected changes to the project schedule that threatened to slow down the progress, especially when the researcher had to create prototypes for monolith and microservice applications in order to experiment with certain system metrics and obtain required numerical data for analysis.

**6.5 Social, Legal, Ethical and Professional Considerations**

To comply with the above, the researcher was made to read, answer all relevant questions and agree to abide by the extant rules guiding research work. Moreover, the research was certified low risk ethics approval.

More, the researcher adheres to the ethical principles of Association of Computing Machinery (ACM) code of ethics and professional conduct (ACM, 2020), especially the following:

**General Ethical Principles**:

1.1 **Be honest and trustworthy**:

being transparent and providing full disclosure of all system capabilities, limitations, and potential problems to the appropriate parties concerned.

1.4 **Be fair and take action not to discriminate**:

The researcher takes this principle into account when comparing monolith and microservice architectures.

1.5 **Respect the work required to produce new ideas, inventions, creative works, & computing artefacts**:

The researcher kept this principle into account and cites relevant sources whose work is used, using appropriate referencing system.

1.6 **Respect privacy and confidentiality**:

This research work did not use interviews, questionnaires and primary data that require personal information.

# Chapter 7

# Conclusion

This thesis set out to review the scope and impact of migrating from monolith architecture to microservices event-driven architecture, how agile methodology plays a role, and the application of bi-modal IT strategy. The study formulated research objectives which were the focus throughout the literature review. Two research questions were set and to answer these, web application prototypes were developed for monolith and microservices architectures and deployed on localhost and docker respectively. Open source metric tools were integrated to measure performance metrics such as response time, and error rate of the two architectures and data was collected and analysed. More, industrial cost of maintenance of the two architectures was also obtained and analysed.

## Achievements

In addressing the first research question: how to migrate to microservices without disrupting ongoing business operations on the system, three possible methods were found and detailed in chapter 5.1, the analysis and discussions chapter. The experience from creating prototypes of monolith and microservices applications showed that converting a monolith to microservice involves converting some modules of the application to independent and separate microservices with their own individual databases. This reduces the number of communications channels and improves performance of the system and ensure better user experience and this was in agreement with the findings of (Strljic et al., 2019). The second method is the bi-modal IT strategy where an organization can maintain the monolith application while building the microservices application at the same time. The third method is converting the monolith application itself, but by migrating the non-customer-oriented functions first, as these hardly affect the operation of the system if problem arises.

To answer research question 2: the scope and impact of migrating to microservices, findings on response time show that if the application is not intended to grow in size and there are no increasing user request calls, a monolith application is best as it outperforms microservices. However, if the application is expected to grow in size and handle growing user requests, then a microservices application is the best option. Analysing error rate also show that microservices better fault tolerance than a monolith. Microservices also allows companies to reduce their infrastructure costs and thus easier to maintain. Finally, microservices enable large applications to be developed as a set of small applications that can be independently implemented and operated by separate teams, thus promoting the application of agile frameworks in the organisations.

# Future Works

The prototypes developed for this experimental study were basic in nature and thus might have affected the accuracy and reliability of the metric results to some extent. For future work, applications with more parts, complex functions, features and calculations hosted on the cloud might potentially produce far reliable results for both monoliths and microservices.

Also, most studies on monoliths and microservices are limited to web and desktop applications. Extending this to investigate the performance differences on mobile applications is highly encouraged.

# Chapter 8

# Student Reflection

8.1 **Personal Performance**

This project work has been a challenge, considering that the topic of investigation has been germane to me, without having prior knowledge and experience in monoliths and microservices. Having recently started a part-time as a web developer with a task to develop an e-commerce application from monoliths to microservices, it was a challenge to venture into a new field, and the opportunity to conduct a research study to investigate migrating from monoliths to microservices has been quite rewarding. In the process I have created prototypes to represent the two architectures, learnt how to experiment and measure system metrics and to optimise them. Further, I learned more about cloud services, pricing, agile frameworks, and project management using management tools.

8.2 **Problems Encountered and Resolution**

As an amateur researcher who is investigating a topic with little previous experience, deciding on the project topic, research methodology and how to conduct thorough investigation has certainly been a great challenge. I had to work on the project topic several times to get it narrowed, spent considerable days looking for available industrial data for investigation, which was difficult to find, and with guidance from the supervisor, had to create own prototypes to experiment and obtain the data for analysis.

8.3 **Lessons Learnt**

The following lessons have been learned:

- Research project is like a full-time work and requires dedication and commitment

- The supervisor will guide and direct on what need to be done, but the student has to cooperate and be self-motivated
- The work can be frustrating, so having a topic one is interested in, with benefit in sight is very important to keep one going when faced with discouragements
- Research work helps us learn new skills and improve on what is already known. I have learned from the knowledge of others on the topic area and added my little discoveries to it.
- Most importantly, I have gained valuable industrial experience on monolith and microservices architecture that will help in current work and future prospects.

## 8.4 **What could have been done better or differently**

The experience would have been quite interesting and enriching, with much better results if I had the opportunity to visit some IT companies, like Gartner, Amazon Web Services or Google, and experimented with large systems. It also could have provided an opportunity to interview and gain perspectives in real time.

# Bibliography and References

# Appendix A – Project Presentation

**Reviewing the Scope and Impact of Implementing a Modernised IT Event-Driven Architecture from Traditional Architecture using Agile Frameworks**

**A Case study of Bi-modal operational strategy**

Sunday Ubur

---

**Problem Statement**

Enterprises and companies are using Traditional Legacy Architecture to maintain their IT presence.

Considered to have Low Complexity, and Easy to Deploy

*However, this architecture has the following problems:*
- Tightly-coupled modules
- Not suitable for Agile teams, which is the standard today
- It costs an organization nearly 80% of IT budget annually to maintain

Microservices Event-driven architecture (EDA) is the solution, but enterprises and companies are reluctant to migrate as they still think EDA is a hype, and they are not aware of the migration to EDA process, its scope and impact to their businesses.

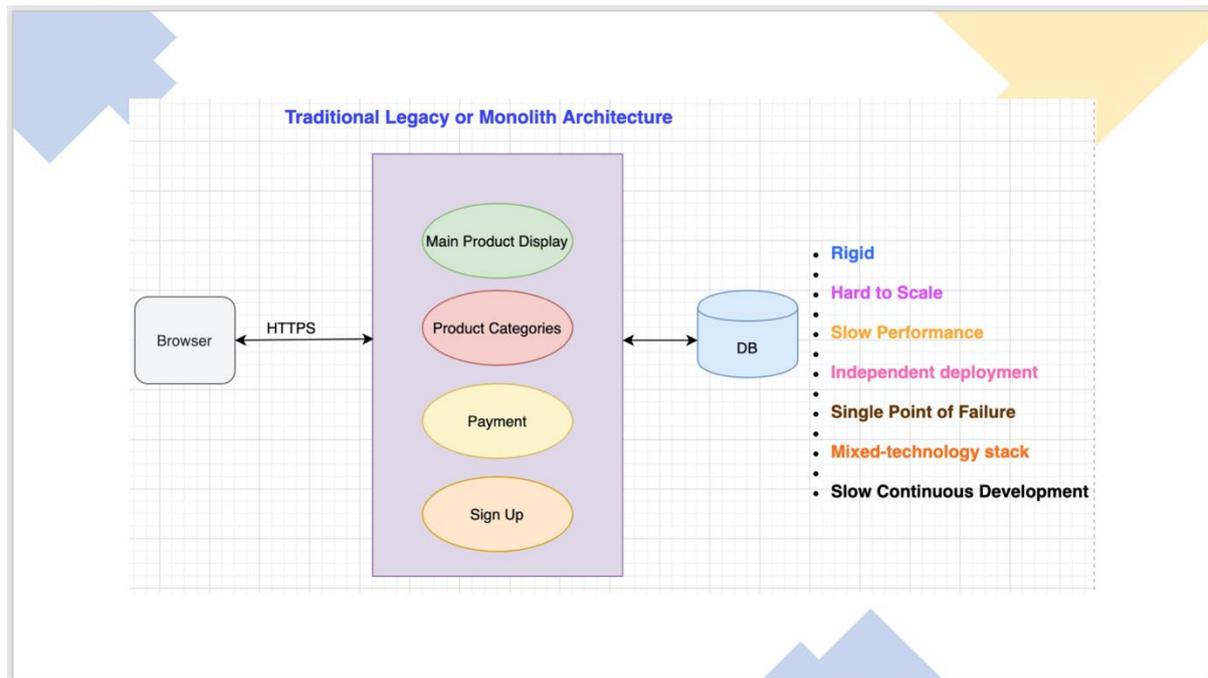

## Objectives of the Study

- A review of Traditional Legacy Architecture and cost of IT maintenance, upgrade and modernisation
- A review of Market adoption of Event-driven Architectures, Trends, Failures and Successes: A Case Study of Netflix
- Analysis of Agile Frameworks in the Transformation from Traditional Legacy Architecture to Event-driven Architecture
- Benefits of Agile Adoption in IT Modernisation when implementing Event-driven Architecture to replace Monolithic Cyber Complex and Enterprise Systems
- Operationalisation of Event-driven architecture using Bi-Modal Strategy in decommissioning systems with Traditional Legacy Architecture

## Progress Made

1. Completed Literature Review:
2. Developed a traditional monolith application prototype
3. Developing a microservice event-driven application prototype
4. Integrated Dropwizard metrics – an open source library for measuring the performance of the applications for the purpose of obtaining real time data to analyse the two architectures.

## Original Project Plan Vs Current Plan

The original project plan was to use available data related to the system metrics to analyse the architectures from dataset websites. However, such data do not exist, so the researcher has resorted to create basic prototypes to obtain required data.

As a result, the work has been behind schedule as regards thorough analysing research findings and conclusions.

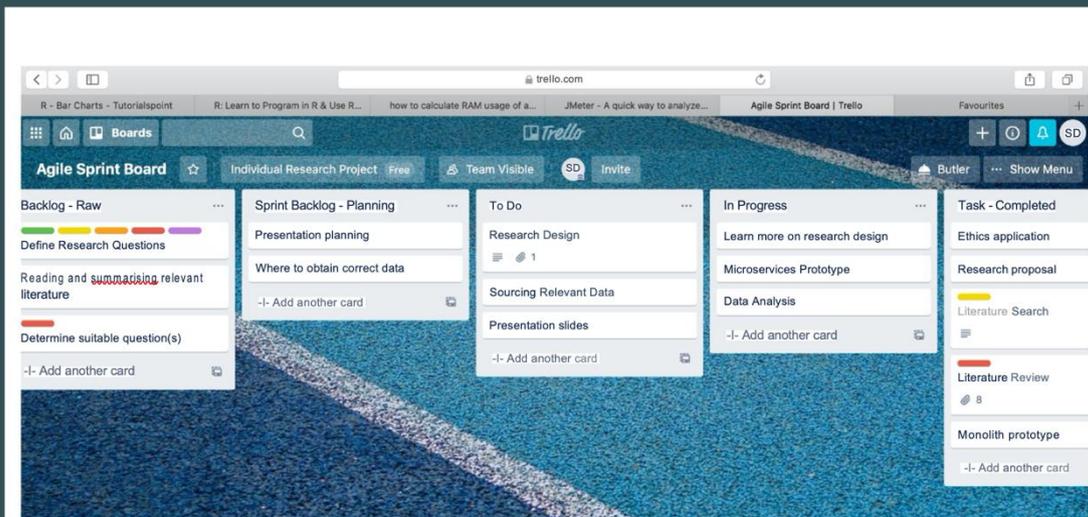

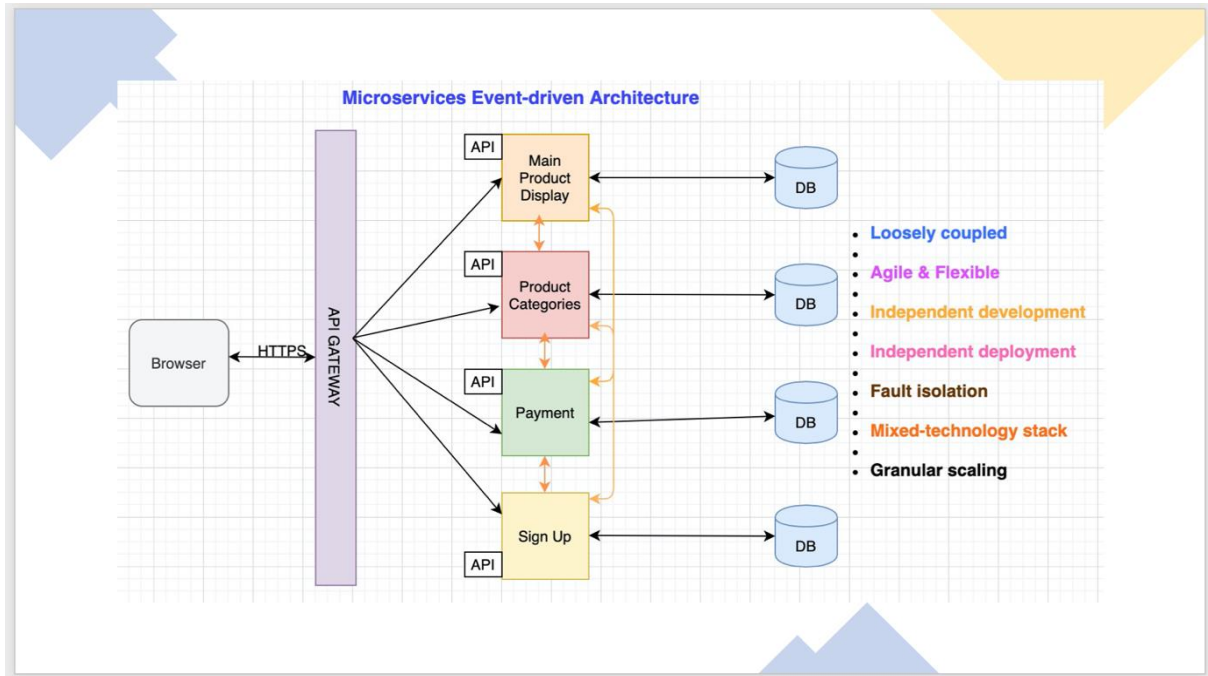

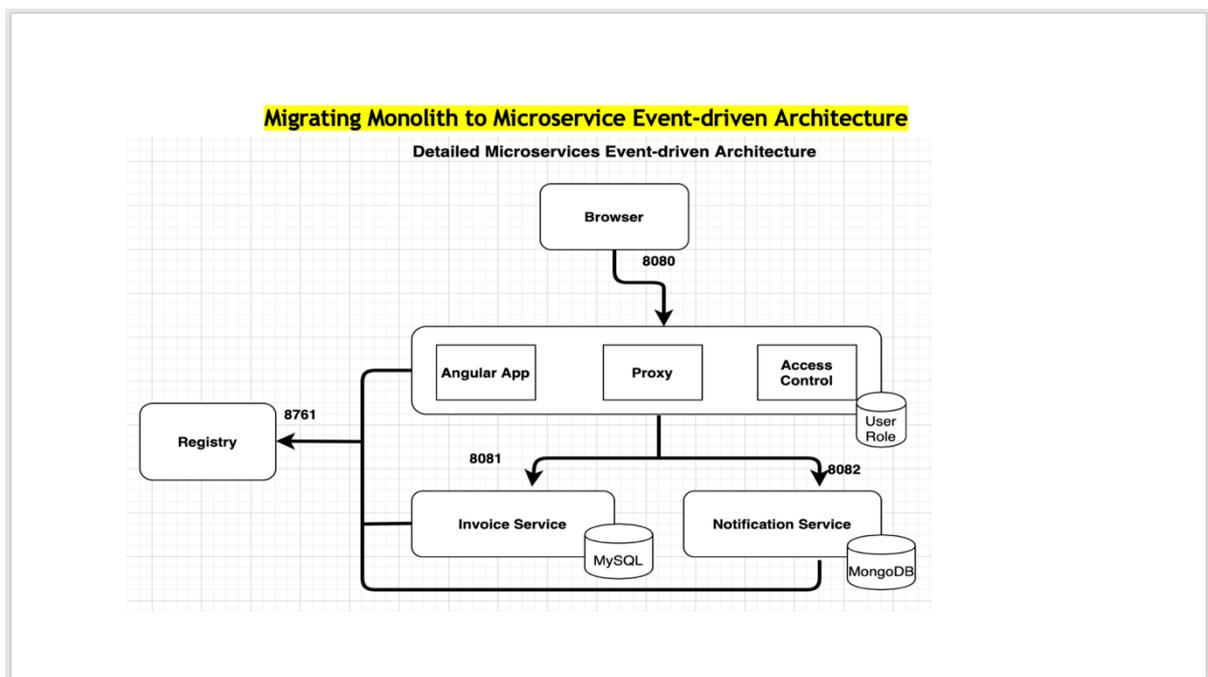

### Initial Data Analysis

In comparing Traditional monolith and Microservices Event-drive architectures, the following metrics was obtained from the experiment with the prototypes:

| Metrics | Monoliths | Microservices |
|---|---|---|
| Response Time (10) users | 283.7 milliseconds | 322.9 milliseconds |
| Response Time (1100) users | 285.33 milliseconds | 210.0 milliseconds |
| Line Response Time (10) | 308 milliseconds | 403 milliseconds |
| Line Response Time (1100) | 354.33 milliseconds | 275.33 milliseconds |
| Error Rate | 62% | 17% |
| Cost per month | $403.20 | $390.96 |
| Communication channels | High | Low |
| JVM Garbage Collection | Mean: 35.667 | Mean: 44.333 |
| Http request in millisecond | Code 200: Mean= 211.33 | Code 200: Mean= 155.91 |
|  | Code 401: mean= 335.66 | Code 401: mean= 240.24 |

## Findings

1. To migrate monolith to microservices without disrupting ongoing business functions, (Sharma, 2017) recommends migrating the non-customer-oriented applications first to microservices, then migrate the customer-oriented functions later.
2. From the metrics data, Microservices has faster response time and better fault tolerance than monolith: it makes better use of system resources
3. Microservices are developed as independent services and supports agility and collaboration. Monoliths don't
4. Cost of maintenance: Monoliths incur higher maintenance costs than microservices.

## Appendix B – Certificate of Ethical Approval

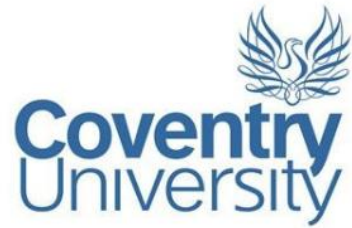

**Certificate of Ethical Approval**

Applicant:

    Sunday Ubur

Project Title:

    Event-Driven Architecture on Enterprise Transactional Systems:
    Comparing the efficiency of traditional software architecture and event-driven architecture

This is to certify that the above named applicant has completed the Coventry University Ethical Approval process and their project has been confirmed and approved as Low Risk

Date of approval:

    08 June 2020

Project Reference Number:

    P107910

## Appendix C- Interim Progress Report and Meeting Records

**Meetings with Supervisor**

20 May 2020　　12:48

**19 May 2020**
First meeting:
Supervisor described the project brief with me using concrete examples of his own project, and that of some students he supervised on previous occasions.

-proposal
Provide a clear outline of the research question
What evidence is there to prove that this is actually a problem?
Intended users, how do they benefit
What is there going to be at the end of it?
Primary research plan to get user feedback
Your job is to answer a question, not to create a piece of software.

What methodology am I going to use?
What tools am I going to use?
What am I going to find out? How am I going to find out?

What data to collect, how to analyse that data?

What statistics do I need to demonstrate that this works?

The software to create should just be a prototype, not a polished work.

Starting point: Find an area that you are interested in - read as much as you can about everything in it that you can find

You are looking for answers to questions that you have.

You can do what others have done, but do it in a different way. (never recreate the wheel;
- Bring in something small; something new; take a small step on top of the next people.
Recommendation:
Make small notes as you read journals on the area you are interested in.

- Summaries
- -quotes
- Reference sources

The research is about two things:
1. Understand the area you're interested in
2. Gather evidences to help you make decisions later
3. Aim for reputable and current sources
4. Read chapter of books that are relevant
5. Conference materials
6. Journals from ACM and IEEE Databases
7. Use google scholar and Wikipedia
8. Keep a copy of every material you download to use

-Recommended to use Zotero software to aid in research
-Use Microsoft word referencing tool to make referencing and citations easier

Project scope: Don't make it too broad or too narrow.

-Make sure you are researching in an area you have interest in. This will ensure you enjoy the whole experience.
-But look for something broad and general to start with

**Changes in Project Topic**

Final Choice of Project Topic

Sunday Ubur
Fri 29/05/2020 13:48
To: Simon Billings

Sir, greetings and hope you're doing well.

Kindly, bear with me for having to change project topic again.

After careful thought and discussing with my employer I have decided to finally settle on:

Impact of event driven architecture on enterprise transactional systems

This research work will be beneficial to the software development discipline, the academia, the business world and my employer of whom I am working on overcoming the current challenge they face with their current system which is tightly coupled. And personally, it will help strengthen my career as software engineer.

Amazon Ware Services defines event driven architecture as "an architecture that uses events to trigger and communicate between decoupled services and is common in modern applications built with microservices."

Right now I am expecting your approval then I will start working on sources for primary research. Currently I am reading articles and conference papers on it on the ACM and IEEE sites.

Thank you for your kind consideration.

**Simon Billings**
Fri 29/05/2020 14:00
To: Sunday Ubur

You'll need to narrow it down a lot, but feel free to start reading.

Simon

…

**Project Proposal Submitted:**

**Research Proposal:**

**Event-Driven Architecture on Enterprise Transactional Systems:** Assessing the efficiency of event-driven architecture over the traditional architecture

**Student ID:**     8460552

**Course Code:**   7048CEM

**Supervisor:**    Simon Billings

**Folllw-up Questions**

**Simon Billings**
Tue 16/06/2020 23:25
To: Sunday Ubur

I think the topic could be good, it needs narrowing down quickly to give a clear idea what you want to discover with your research. If you want to build a system then you want to know what decisions you'll have to make in implementing it. Hopefully you'll find that you're not sure, from the secondary research, which of several alternatives is the best decision to make. If that is the case then investigating that through primary research is the project and you can start to plan how you test the different alternatives and compare them.

If you find that there is a clear best way to approach it then see if you can apply that approach in a context other than that which it's already been tested. You can then compare your experience applying it with what you've read, ideally by gathering the same kind of data they've gathered in what you've read so you can do the same kind of analysis and compare them.

I don't think you'll need an alternate topic, but it's worth giving a little thought to it in case neither of the two things above seem possible.

Simon

**Final Project Proposal**

1. **Topic of Investigation**:
   Reviewing the Scope and Impact of Implementing a Modernised IT Event-Driven Architecture from Traditional Architecture using Agile Frameworks – A Case study of Bi-modal operational strategy

   1.1 **Problem Statement**:

   Many business enterprises and companies that adopt IT as critical part of their daily operations still depend on traditional or legacy architecture software although most of the components are becoming obsolete and incur about 80 percent of the IT department's operational costs (Newman, 2017). Also, such businesses and companies find it difficult to apply modern IT practices such as Agile frameworks as it is difficult for a team of developers to work collaboratively on a traditional, legacy or monolith architected software, thereby making maintenance even more difficult. Salescache, an online discounts shop is one example of several companies whose enterprise system was build using the monolith approach. It resulted in the system being tightly coupled and so making it difficult to maintain, and the entire system needing to be rebuilt using the event-driven architecture framework. However, findings show that many companies are sceptical in migrating to the event-driven microservices, not because they don't embrace the new technology and the benefit agility offers, but because they don't understand how to migrate while ensuring little or no disruption to the old system (Purcell, 2017). There is also the problem of financial aspects of migrating to microservices. Managed services in public cloud platforms have very different pricing models, and the choice of a microservice runtime or data storage solution can have a significant impact on the cost-effectiveness of running a microservice system in the cloud. A price study comparing the costs for running microservice with different cloud platform managed services would help architects and enterprise companies in choosing architectural patterns also with cost-effectiveness in mind. These issues inspired this research work to attempt and investigate the efficiency of event-driven architecture compared to the traditional architecture, analyse the scope and impact of migrating, and the benefits agile adoption will have on such enterprises.

1.2 **Research Objectives**:
- vi. A review of Traditional Legacy Architecture and cost of IT maintenance, upgrade and modernisation
- vii. A review of Market adoption of Event-driven Architectures, Trends, Failures and Successes. A case study of Netflix
- viii. Analysis of Agile Frameworks in the Transformation from Traditional Legacy Architecture to Event-driven Architecture
- ix. Benefits of Agile Adoption in IT Modernisation when implementing Event-driven Architecture to replace Monolithic Cyber Complex and Enterprise Systems
- x. Operationalisation of Event-driven architecture using Bi-Model Strategy in decommissioning systems with Traditional Legacy Architecture

1.3 **Intended Users**:

i. The academic community: At the end of this project, we expect to have more in-depth knowledge of various event-driven architectures currently available and determine the preferred architecture for enterprise systems.

ii. Software developers aiming to use the event-driven architecture in their practices: Encourage painless use of agility as collaborating teams will be dealing with separate modules that are highly independent and loosely coupled.

iii. Salescache: the report may help the company know how to maintain their system going forward, as well as make the best choice of architecture in their future developments.

iv. Career progression

2. **System Requirements, Project Deliverables and Final Project Outcome**

   **System Requirement**

   i. RapidMiner studio for the impact-cost analysis of different cloud service platforms
   ii. Microsoft Excel as database for data to be analysed with rapidMiner
   iii. VisualParadigm to design some architectural diagrams

   **Project Deliverables**

   i. Gathering data, processing the data, analysing and discussions of findings of the impact-cost analysis of financial involvement using different cloud service platforms
   ii. Recommending the best cost-effective platforms for enterprise adoption

   **Final Project Outcome**

   The outcome will be the analysis results, benefits of agile adoption using event-driven architecture for enterprises, and recommending best practices.

3. **Primary Research Plan**
    i. Data of cloud migration plans obtained from providers websites
    ii. Factual data from Salescache

**Project Timeline Schedule**

|      | Activity | Duration |
|------|----------|----------|
| i.   | Project Topic Definition | 26$^{th}$ May – 18$^{nd}$ June |
| ii.  | Secondary data collection and processing | 27$^{th}$ May – 25$^{th}$ June |
| iii. | Primary Data cleaning, storage and processing | 26$^{th}$ June- 5$^{th}$ July |
| iv.  | Discussion of findings | 6$^{th}$ July – 9$^{th}$ July |
| v.   | Working on CW2 | 9$^{th}$ July – 16$^{th}$ July |
| vi.  | Documentation | 14$^{th}$ July – 10$^{th}$ August |

## Meeting on Research Methodology

**Simon Billings**
Mon 22/06/2020 21:13
To: Sunday Ubur

You can use secondary data, but you should be conducting an original analysis of that data to draw your own conclusions - it is not enough to rely on conclusions drawn by others, ie. secondary research.

Simon

...

**Sunday Ubur**
Sir, Thank you. That's fair enough, and well noted. Regards, Sunday Sent from Outlook Mobile      Mon 22/06/2020 21:23

**Simon Billings**
Mon 29/06/2020 13:16
To: Sunday Ubur

You'll want to demonstrate that you've found a source a data, and show the progress you've made in cleansing, analysing, and interpreting it. You might find writing software to parse, cleanse, and analyse the data is a good use of your time and a good demonstration of your technical skills. It's pretty common for researchers to write crude code to automate data processing tasks like this - I write code to automate bulk uploading and marking processes where I can.

Simon

...

**Project**

**Sunday Ubur**
Thu 09/07/2020 17:07
To: Simon Billings

📄 Work in Progress.docx
668 KB

Sir,
Good day and thank you for your email.

Kindly, can I share the work I have done so far, so you will go through and have a sense where it's moving?

Actually, I intend to work with data to analyse scope and impact, and the cost of maintaining a traditional or monolith architecture compared to micro services, and the difference between their performance, so kindly do go through the work and offer a way forward to ensure I'm heading in the right direction.

More so, I am wonder if I can use SPSS.

If a software demonstration is required to describe Microservices, remember I'm working on migrating a system from monolith to microservices. But, because I understand this is a research project and not software development project as you said, I decide not to include the work I'm doing on the website.

Please find attached my ongoing work, and awaiting your kind response.

Regards,

**Simon Billings**
Thu 09/07/2020 17:28
To: Sunday Ubur

Hi Sunday,

I skimmed through the litterature review, which seems reasonable, but there's not much else there for me to comment on: with respect to the methodology for example. Assuming you've suitable data to analyse to answer the questions posed yourself, rather than relying on the answers of others, there shouldn't be an issue. Given that you're talking about using SPSS I assume you have data that you can conduct your own analysis with. You can use any software or tools you want, including SPSS, but remember you want to demonstrate technical skills where you can. If you're not creating software as part of an experiment, writing code to conduct the analysis yourself is one way to show these skills. If you have other ways to demonstrate them in the project then SPSS is a perfectly reasonable tool to use for the analysis, you might also consider R if you're looking for something more flexible to connect several processes together.

Best regards,

Simon

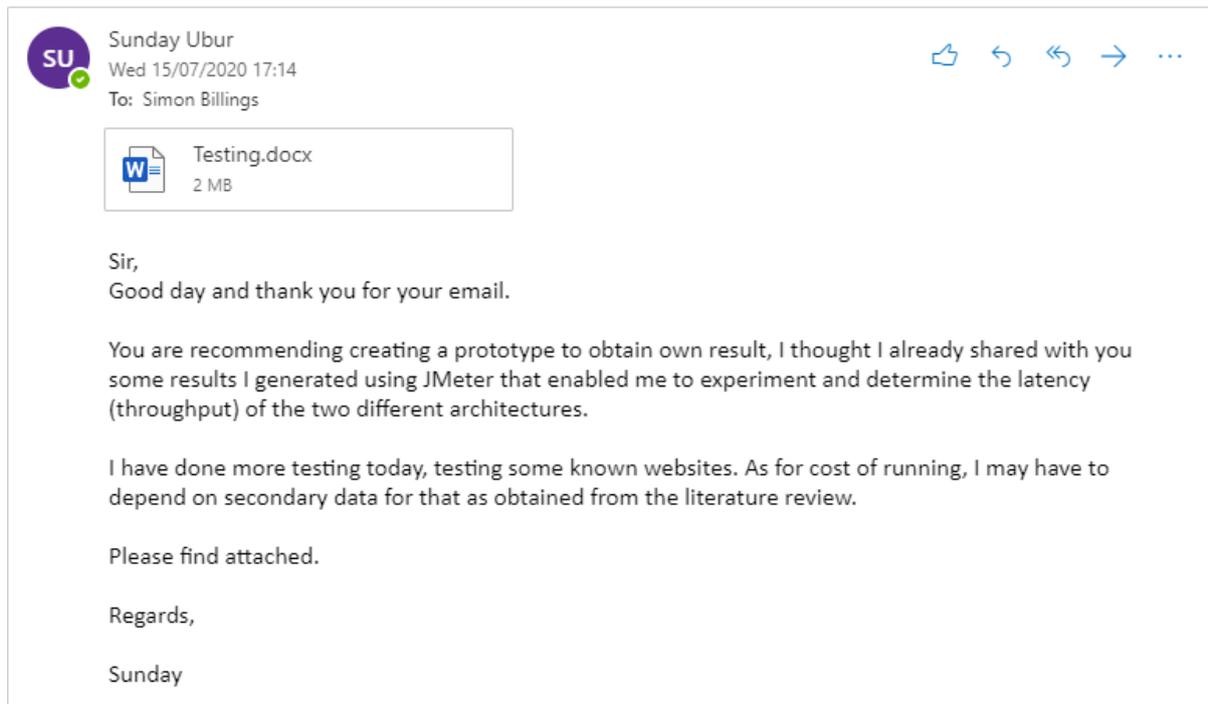

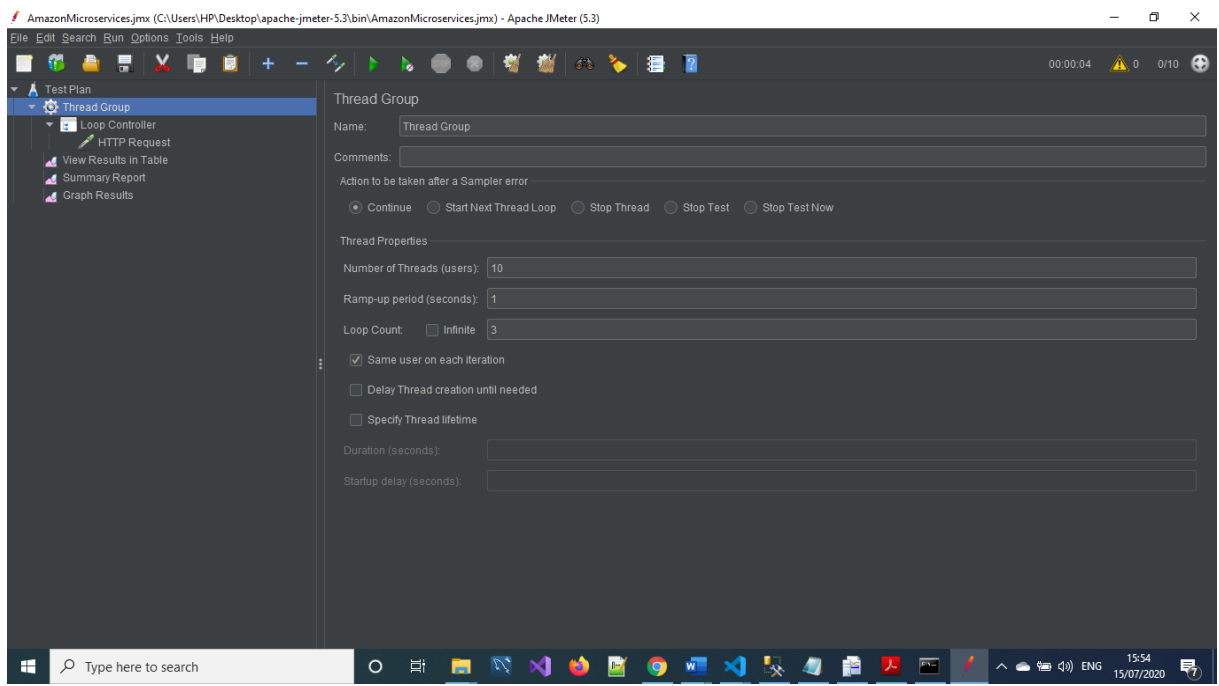

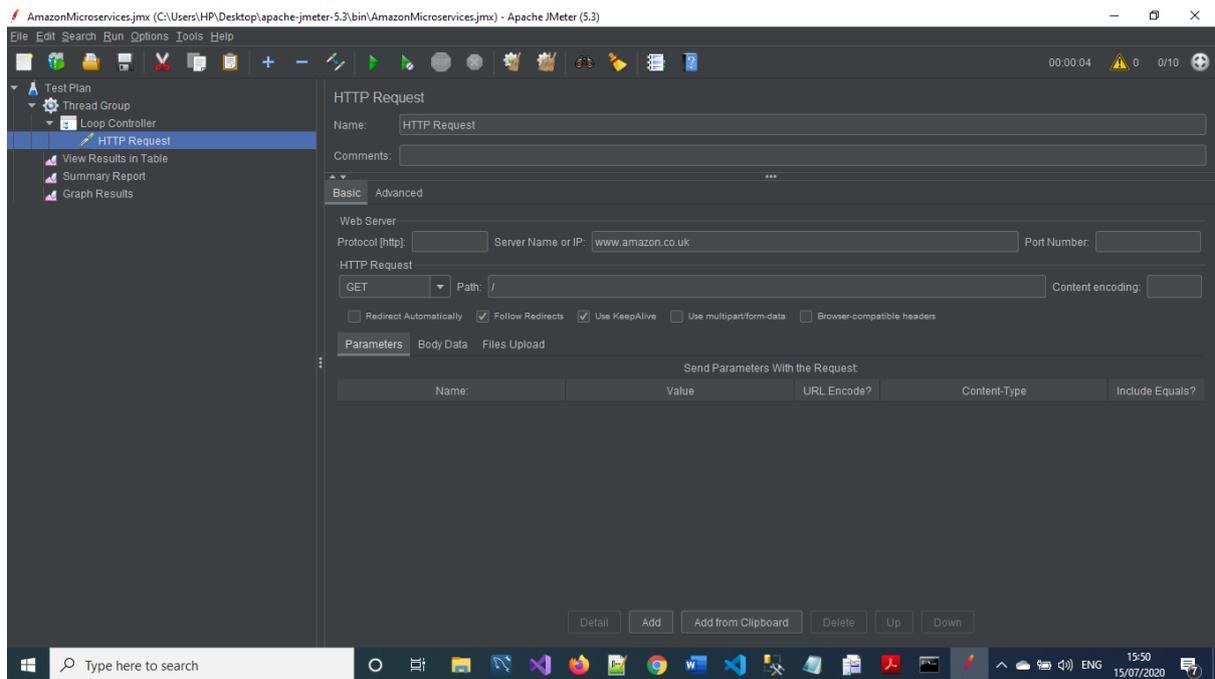

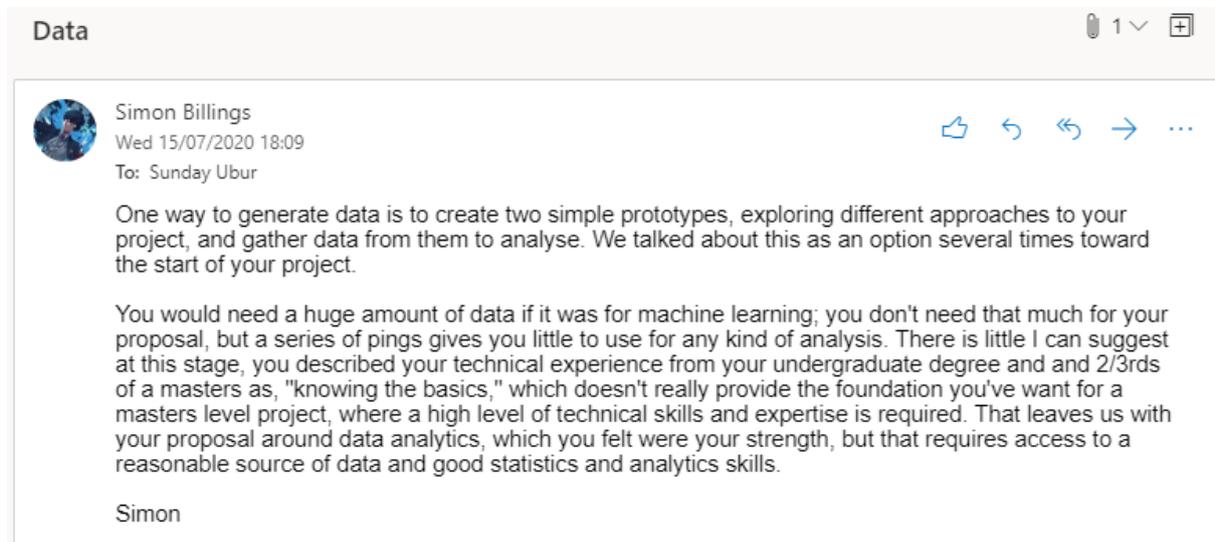

**Data from Prototypes**

**Sunday Ubur**
Mon 20/07/2020 10:41
To: Simon Billings

📎 Microservice Gateway Metrics... 78 KB
📎 Microservice Invoice Metrics.... 73 KB

Show all 4 attachments (292 KB)   Download all   Save all to OneDrive - Coventry University

Sir, good morning and hope you're doing well.

I wish to show you the data obtained from the monolith and the microservices developed.

The monolith is just a single application, while the microservice has a main application: Gateway that has the user interface, and 2 independent microservices: notification and invoice, as well as a registry that has an open source application - Dropwizard for measuring the metrics, and the microservice application hosted on Docker.

I will extract what I need from the metrics as not everything is needed.

Meanwhile, I wonder if my boss will be needed to participate in the presentation. I have informed him and he said he might try. His email is: michael.orji1@gmail.com

Do have a nice day Sir.

---

**Data from Prototypes**

**Simon Billings**
Mon 20/07/2020 11:04
To: Sunday Ubur

That's great and it gives you a much better pool of data to use for analysis 😀.

We shouldn't need your boss for the presentation, it's primarily a chance to get feedback and a means of academic assessment. But there's nothing to stop you repeating the presentation for your boss and colleagues at their convenience, they'll probably have different questions about the project than we would.

Simon

...

**Sunday Ubur**
Mon 20/07/2020 11:31
To: Simon Billings

Thank you Sir.

I'm very grateful, and so much motivated.

Regards,

Sunday